\documentclass[twocolumn,twoside]{IEEEtran}                 

\usepackage{subfigure}
\usepackage{bm}
\usepackage{graphicx}
\usepackage{amsmath, amsthm, amsfonts, amssymb, amsbsy}
\usepackage{setspace}
\usepackage{graphicx}
\usepackage{pstricks,arydshln}
\usepackage{multirow}
\usepackage{booktabs}
\usepackage{stfloats}
\usepackage{algorithmic,amscd,textcomp,amssymb,epsfig,graphics,epsf,color,amsmath,balance,algorithm,cite}
\usepackage{caption}
\usepackage{cite}
\usepackage{epstopdf}
\usepackage{dsfont}
\usepackage{color}
\usepackage{algorithm}
\usepackage{algorithmic}

\begin{document}
\title{A Promising Technology for 6G Wireless Networks: \\Intelligent Reflecting Surface}

\author{Wen-Xuan Long,\,\IEEEmembership{Student Member,\,IEEE,}\,Rui~Chen,~\IEEEmembership{Member,~IEEE,}\,Marco Moretti,~\IEEEmembership{Member,~IEEE},\\Wei Zhang, \IEEEmembership{Fellow,~IEEE} and Jiandong Li, \IEEEmembership{Fellow,~IEEE}
\thanks{This work was supported in part by the open research fund of National Mobile Communications Research Laboratory, Southeast University under Grant number 2021D04, the Fundamental Research Funds for the Central Universities, and the Innovation Fund of Xidian University.}
\thanks{W.-X. Long is with the State Key Laboratory of ISN, Xidian University, Xi'an 710071, China, and also with the University of Pisa, Dipartimento di Ingegneria dell'Informazione, Italy (e-mail: wxlong@stu.xidian.edu.cn).}
\thanks{R. Chen is with the State Key Laboratory of ISN, Xidian University, Xi'an 710071, China, and also with the National Mobile Communications Research Laboratory, Southeast University, Nanjing 210018, China (e-mail: rchen@xidian.edu.cn).}
\thanks{M. Moretti is with the University of Pisa, Dipartimento di Ingegneria dell'Informazione, Italy (e-mail: marco.moretti@iet.unipi.it).}
\thanks{W. Zhang is with the School of Electrical Engineering and Telecommunications, The University of New South Wales, Australia, (e-mail:
w.zhang@unsw.edu.au).}
\thanks{J. Li is with the State Key Laboratory of ISN, Xidian University, Xi'an 710071, China (e-mail: jdli@mail.xidian.edu.cn).}%
}

\maketitle

\begin{abstract}
The intelligent information society, which is highly digitized, intelligence inspired and globally data driven, will be deployed in the next decade. The next 6G wireless communication networks are the key to achieve this grand blueprint, which is expected to connect everything, provide full dimensional wireless coverage and integrate all functions to support full-vertical applications. Recent research reveals that intelligent reflecting surface (IRS) with wireless environment control capability is a promising technology for 6G networks. Specifically, IRS can intelligently control the wavefront, e.g., the phase, amplitude, frequency, and even polarization by massive tunable elements, thus achieving fine-grained 3-D passive beamforming. In this paper, we first give a blueprint of the next 6G networks including the vision, typical scenarios and key performance indicators (KPIs). Then, we provide an overview of IRS including the new signal model, hardware architecture and competitive advantages in 6G networks. Besides, we discuss the potential application of IRS in the connectivity of 6G networks in detail, including intelligent and controllable wireless environment, ubiquitous connectivity, deep connectivity and holographic connectivity. At last, we summarize the challenges of IRS application and deployment in 6G networks. As a timely review of IRS, our summary will be of interest to both researchers and practitioners engaging in IRS for 6G networks.
\end{abstract}


\section{Introduction}

The fifth (5G) generation wireless communication networks have basically completed the preliminary basic tests, the construction and standardization of hardware facilities, and would be deployed globally from 2020. Compared with the fourth generation (4G) wireless communication networks, 5G networks have achieved revolutionary advancement in data rate, latency, reliability, mobility and large-scale connectivity$^{\rm [1]}$, which can achieve $20$ Gb/s peak data rate, $0.1$ Gb/s user experience data rate, $1$ ms end-to-end latency and support $500$ km/h mobility, $10^6$ devices/$\rm km^2$ connection density, $10$ Mb/s$/\rm m^2$ area traffic capacity. IMT 2020 has been proposed the three main communication scenarios of 5G networks: enhanced mobile broadband (eMBB), large-scale machine type communication (mMTC), and ultra-reliable and low-latency communication-interaction (uRLLC). To improve the communication performance, 5G networks apply various advanced technologies, such as the millimeter wave (mmWave), massive multiple-input multiple-output (MIMO) and ultra-dense network (UDN)$^{\rm [2]}$. However, the key technologies applied in 5G networks, such as UDN and massive MIMO, will bring high hardware costs and huge energy consumption expenditures$^{\rm [3], [4]}$. Therefore, how to reduce hardware costs and energy consumption becomes a key issue that needs to be solved urgently in next-generation communication networks.

With the global commercialization of 5G networks, research institutions at home and abroad have begun to set their sights on the sixth generation (6G) wireless communication networks. In 2017, the European Union launched a three-year research project on the basic 6G technologies$^{\rm [5]}$. In 2018, Finland announced a research program ``6Genesis" aiming at developing, implementing and testing key enabling technologies for 6G networks$^{\rm [6]}$. Besides, the United States and the United Kingdom have invested in some potential techniques for 6G networks such as terahertz (THz)-based communications and quantum technology$^{\rm [7]}$. An official statement from China's Ministry of Industry and Information Technology indicates that China has concentrated on the development of 6G networks.

\begin{figure}[t]
  \centering
  \includegraphics[scale=0.20]{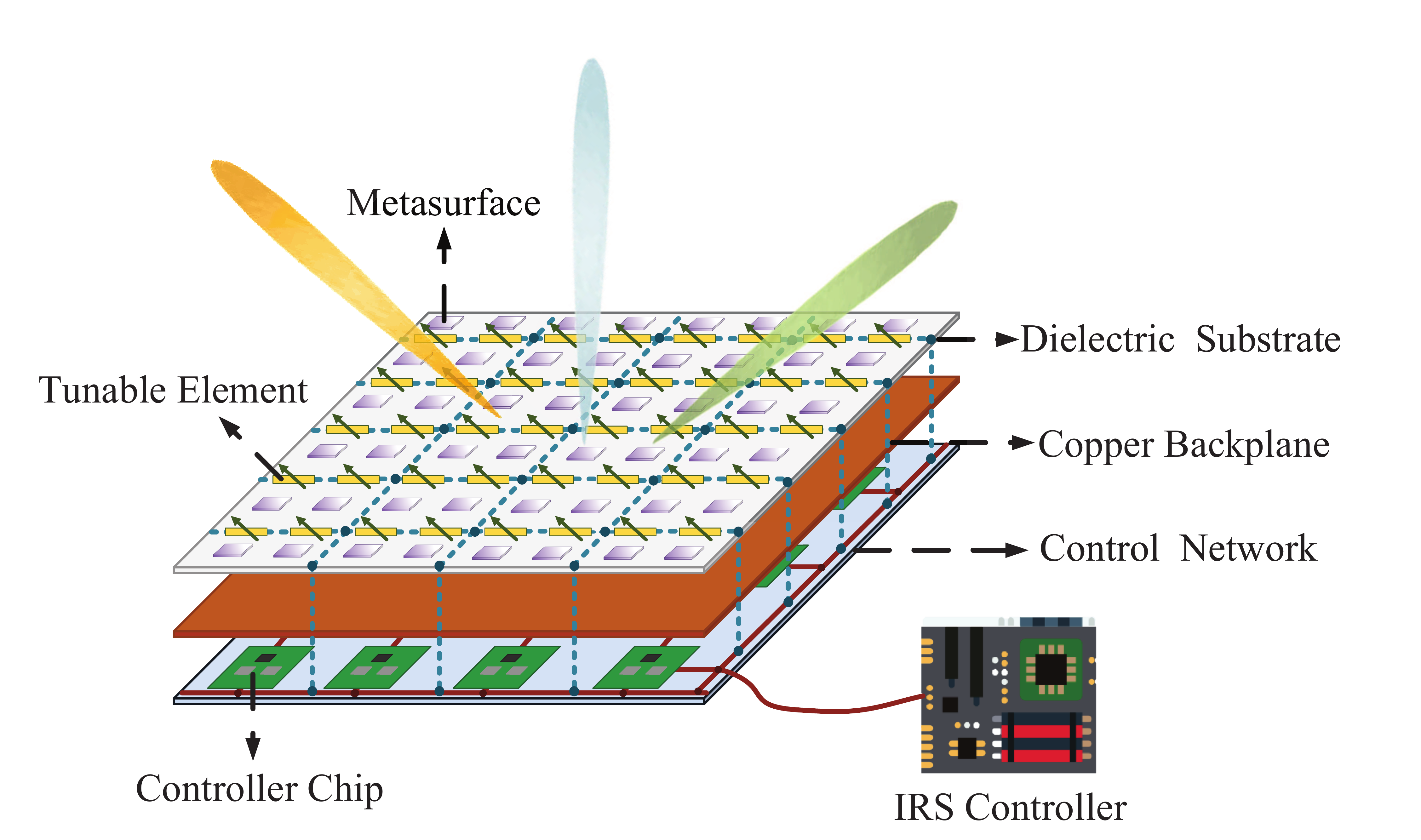}\\
  \caption{The architecture of IRS.}
  \label{Fig1}
\end{figure}

The key drivers of 6G networks result not only from the challenges and performance limits that 5G networks present but also from the growing demand for wireless networks and the technology-driven paradigm shift. The intelligent information society$^{\rm [8]}$ creates core requirements for 6G networks that will lead to the vision of intelligent connectivity, ubiquitous connectivity, deep connectivity and holographic connectivity. Meanwhile, 6G networks will largely enhance and expand application scenarios on the basis of 5G networks. It is reported in Ref. [9] that ubiquitous mobile ultra-broadband (uMUB), ultra-high data density (uHDD) and ultra-high-speed with low-latency communications (uHSLLC) will become the main communication scenarios of 6G networks. The rapid growth of emerging applications has led to a surge in mobile data traffic. Therefore, the key performance indicators (KPIs) for 6G networks should be improved greatly to meet upcoming services. It is reported in Ref. [10] that the peak rate of 6G networks can be $1$-$10$ Tb/s, the connection density can be up to $10^7$ devices/$\rm km^2$ and the energy efficiency is $10$-$100$ times compared to 5G networks. In order to provide satisfying services for the intelligent information society in the future, 6G networks need to further enhance their scalability, flexibility, and efficiency by embracing novel techniques. Like the emergence of many new technologies when the wireless world moves toward 5G networks, the new requirements of 6G networks will influence the main technology trends in its evolution process. In many potential technologies, the intelligent reflecting surface (IRS), benefited from the breakthrough in the manufacture of programmable meta-material, is conjectured as a crucial enabling technology for 6G networks to achieve intelligent radio environments.

In the existing 5G networks, the channel environment is uncontrollable and modeled by probability$^{\rm [11], [12]}$, which is a major limiting factor for wireless communication networks. Specifically, since mmWave frequencies have been allocated to 5G networks, high directivity makes mmWave communications vulnerable to blockage resulting in unstable network performance. To improve the performance of wireless networks, the concept of an intelligent information network with the controllable channel environment is proposed in 6G networks, in which IRS is an essential technology to address the blockage issue and realize uninterrupted
wireless connectivity for mmWave and THz networks$^{\rm [7], [9], [13]-[19]}$. Generally, IRS is a planar antenna array that consists of a large number of low-cost passive reflecting elements, in which each element can intelligently reconfigure the amplitude, phase, frequency and polarization of incident signals in real time and then reflects the specified receiver to enhance signals and suppress interferences. The typical architecture of IRS is shown in Fig. \ref{Fig1}.

\begin{figure*}[tb]
\setlength{\abovecaptionskip}{0.5cm}   
\setlength{\belowcaptionskip}{0.2cm}   
  \centering
  \includegraphics[scale=0.295]{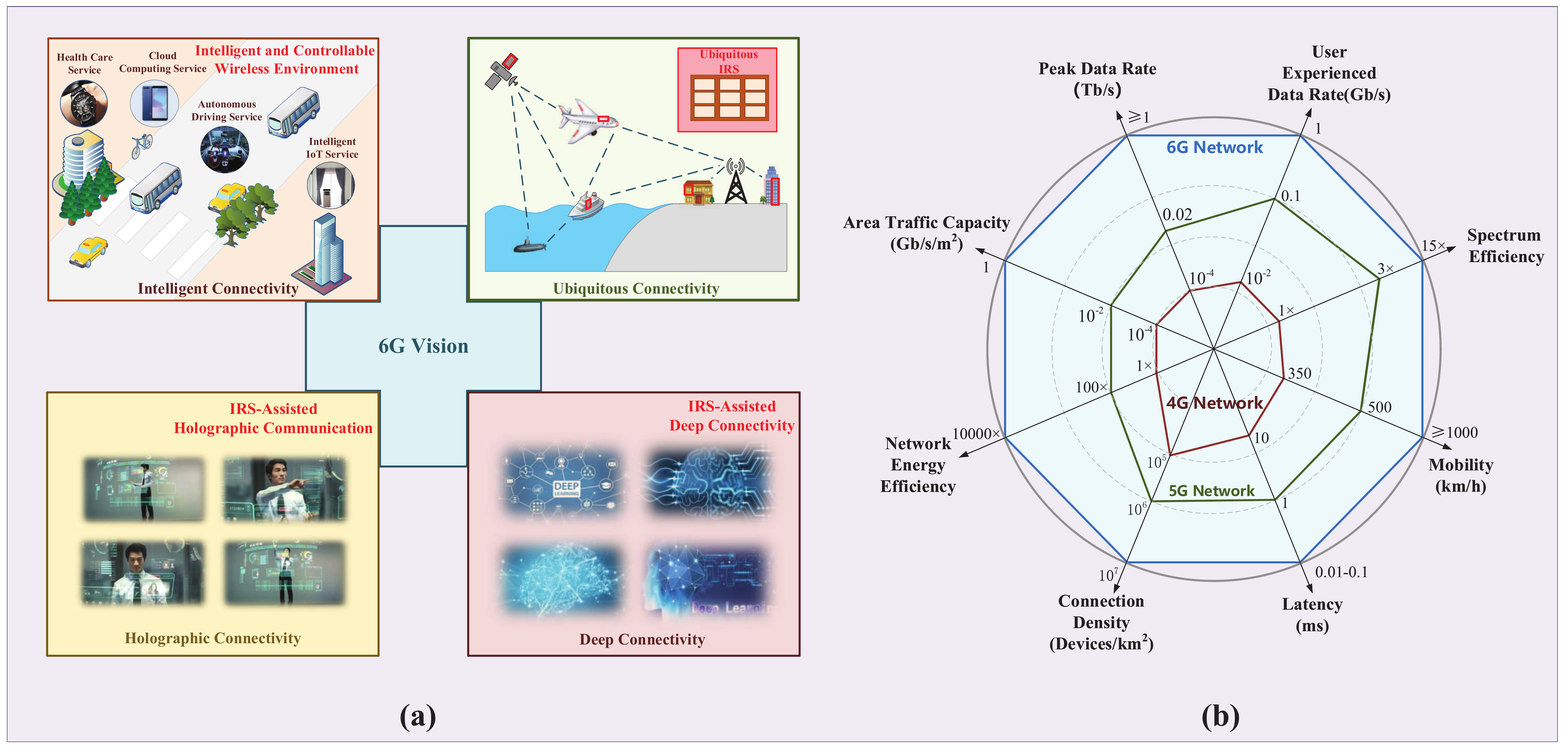}\\
  \caption{The blueprint of 6G networks. (a) The vision of 6G networks with IRS technology, (b) The KPI comparison of 4G, 5G and 6G networks.}
  \label{Fig2}
\end{figure*}

As a new concept, IRS is first proposed in 2017$^{\rm [20]}$ and has attracted wide attention in the wireless communication fields$^{\rm [21]-[30]}$. In 2018, the authors in Ref. [21], [22] prove IRS technology can be applied in wireless data transmission and position estimation for the first time. In the same year, the authors in Ref. [23], [24] propose a novel IRS-assisted wireless communication network, which opens a new wireless communication paradigm. In 2019, the authors in Ref. [25] design and realize the IRS-assisted wireless communication system, which achieves 2.048 Mbps data transfer rate with video streaming transmission over the air. In the same year, the authors in Ref. [26] propose an IRS-based $8$-phase shift-keying (8PSK) wireless transmitter with $8 \times 32$ phase adjustable unit cells, which can achieve $6.144$ Mbps data rate over the air at $4.25$ GHz with less hardware complexity. Besides, the authors in Ref. [27] propose an IRS-based resource allocation method for the downlink multi-user communication system, which are able to provide up to $300\%$ higher energy efficiency in comparison with the use of regular multi-antenna amplify-and-forward relaying. In 2020, The authors in Ref. [28] propose IRS-space shift keying (IRS-SSK) and IRS-spatial modulation (IRS-SM) schemes, which bring the concept of IRS-assisted communications to the realm of index modulation (IM) for the first time. In the same year, the authors in Ref. [29] propose and develop a new type of high-gain yet low-cost IRS that can achieve $21.7$ dBi antenna gain at $2.3$ GHz and $19.1$ dBi antenna gain at $28.5$ GHz. Besides, the authors in Ref. [30] design an IRS-assisted radio-frequency (RF) sensing system for posture recognition, which has $23.5\%$ higher recognition accuracy than the traditional RF sensing system due to the environment control capability of IRS. The above results have laid the foundation for the application of IRS in 6G networks.

It is worth noting that the IRS is different from other related technologies currently employed in wireless networks, such as relaying and backscatter communications$^{\rm [31]}$. Details will be provided in the sequel, but it suffices to say that IRS has the following distinguishable features:
\begin{itemize}
\item
{\bfseries Nearly passive}: IRS composes of a large number of low-cost passive reflecting elements that are only used to reflect signals and do not need to transmit signals. Hence, IRS is nearly passive, and, ideally, does not need any dedicated energy source.
\item
{\bfseries Programmable control}: IRS can control the scattering, reflection, and refraction characteristics of the radio waves by programs, thus overcoming the negative effects of natural wireless propagation. Hence, IRS-assisted wireless communications can intelligently control the wavefront, such as the phase, amplitude, frequency and even polarization, of the impinging signals without the need for complex decoding, encoding, and radio frequency processing operations.
\item
{\bfseries Good compatibility}: IRS can be integrated into the existing communication networks only by changing the network protocol without changing hardware facilities and software of their devices. Meanwhile, IRS has a full-band response, which can work at any operating frequency ideally.
\item
{\bfseries Easy deployment}: IRS is characterized by small size, light weight, conformal geometry and thinner than wavelengths, which is easier to install and dismantle. Hence, IRS can be easily deployed, e.g., on the outside walls of buildings, billboards, ceilings of factories and indoor spaces, human clothing, etc.

\end{itemize}
These unique characteristics make IRS-assisted communications a distinctive technology, which can be regarded as the key to realize the intelligent information society in 6G networks. The application of IRS in 6G networks will be discussed and elaborated in the sequel.

The rest of this paper is summarized as follows. In Section II, we summarize the vision, typical scenarios and KPIs of 6G networks. In Section III, an overview of the IRS technology is provided including the new IRS-assisted system model, the hardware architecture of IRS and competitive advantages in 6G networks. Based on the vision of 6G networks, we propose a general idea of IRS-assisted 6G wireless networks and force on the application of IRS in the connectivity of 6G networks in Section IV, the challenge of IRS application and deployment in 6G networks and the conclusion are summarized in Section V.

\section{6G Networks: Vision, Usage Scenarios and Requirements}

The intelligent information society, which is highly digitized, intelligence inspired and globally data driven, will be deployed in the next decade$^{\rm [8]}$. 6G networks are the key to achieve this grand blueprint, which is expected to offer the connection for everything, full wireless coverage and the integration of all functions so that it can support full vertical applications$^{\rm [10]}$. Hence, 6G networks are required to process a very high volume of data in near real time, with extremely high throughput and low latency. In the following we summarize the vision, typical scenarios and KPIs of 6G networks.

\subsection{Vision}

The development of intelligent information society is the driver for 6G networks that will ubiquitously support high-precision and holographic communications for upcoming intelligent information services to provide a full sensory experience. Overall, the vision of 6G networks can be split into four main points$^{\rm [14], [15]}$:
\begin{itemize}
\item
{\bfseries Intelligent connectivity}: 6G networks will be an autonomous system with human-like intelligence, which will provide multiple ways, such as through voice, eyes and brain waves, to interact with intelligent terminals. Therefore, ``Intelligent connectivity" will require simultaneously: 1) intelligent management should be provided for the complex network, 2) all the related connected devices are intelligent, and 3) the related information services need to be intelligent.
\item
{\bfseries Ubiquitous connectivity}: With the rapid development of aerospace and deep-sea exploration technologies, the active space of human beings and intelligent devices will be expanded greatly, which puts higher requirements on the coverage of communication networks. Therefore, one goal of 6G networks is to achieve ubiquitous connectivity by integrating satellite communications, aerial communications, terrestrial communications and underwater communications to provide global coverage$^{\rm [7], [32]}$.
\item
{\bfseries Deep connectivity}: With the development of intelligent information services, such as the intelligent internet of things (IoT), the types and scenarios of information interaction are becoming more and more complex. There is reason to believe that the information interaction in 6G networks will be greatly expanded in both space and information types. Therefore, ``Deep connectivity" will require simultaneously: 1) the active space of each connected intelligent devices is expanded in depth, and 2) the network itself has the ability of deep sensing, deep learning and deep mind, which is expected to realize the mind-to-mind interaction with intelligent devices.
\item
{\bfseries Holographic connectivity}: With the dramatic increase of augmented reality(AR) and virtual reality (VR) technique, the form of information interaction in the next decade will likely evolve into the high-fidelity VR/AR interaction, and even the holographic information interaction. Therefore, 6G networks will be required to provide ubiquitous high-fidelity VR/AR services and even support holographic communications.
\end{itemize}

In sum, 6G networks will be further developed and strengthened on the basis of the existing 5G networks, and truly realize ubiquitous intelligent interaction between human beings and everything, which is ``wherever you think, everything follows your heart". The vision of 6G networks is illustrated in Fig. \ref{Fig2} (a).

\begin{figure}[tb]
  \centering
  \includegraphics[scale=0.55]{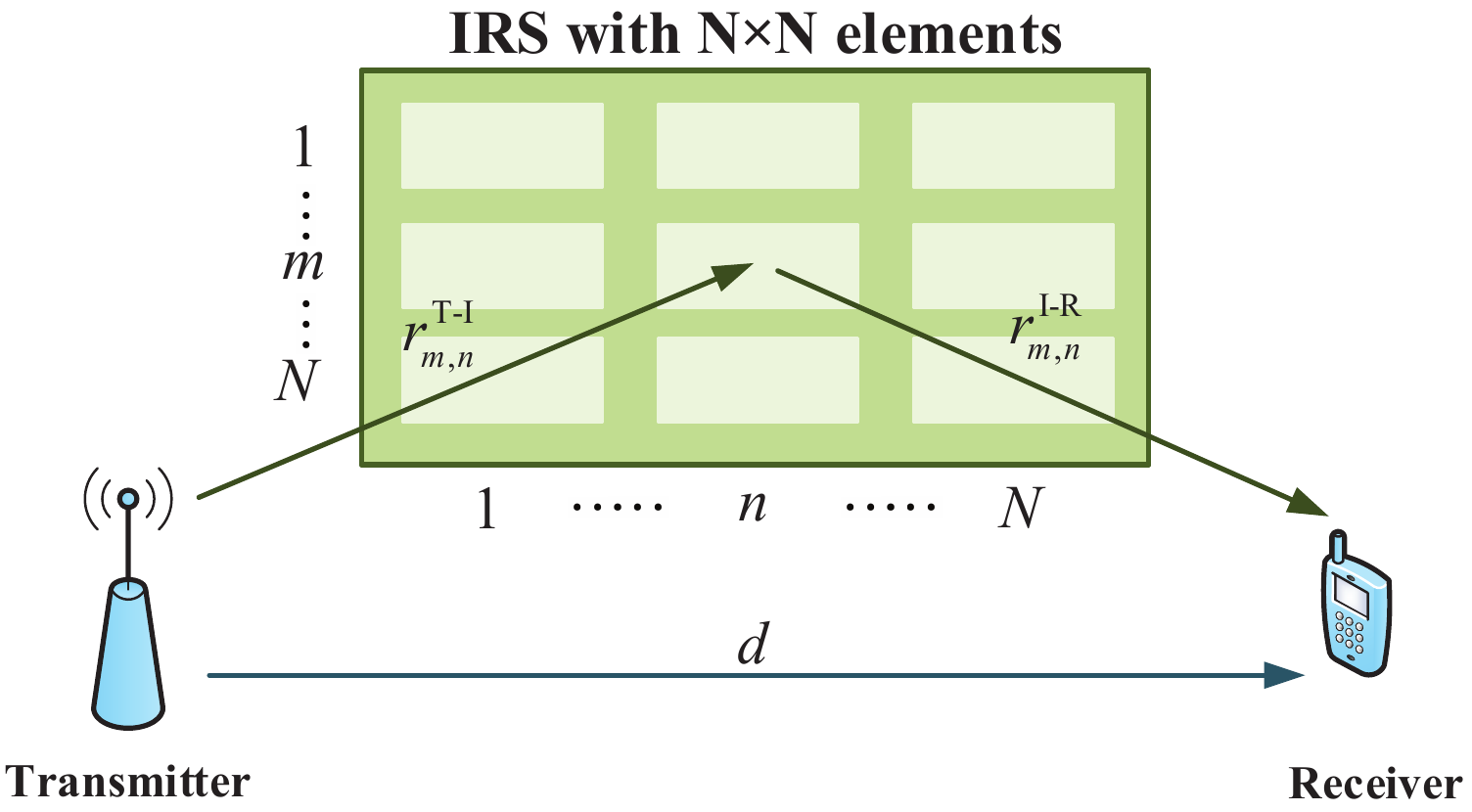}\\
  \caption{The IRS-assisted two-ray channel model.}
  \label{Fig3}
\end{figure}
\subsection{Typical Scenarios}

As all known, 5G networks mainly focus on eMBB, mMTC, and uRLLC, while the goal of 6G networks is to further enhance and extend the application scenarios. In Ref. [9], the typical scenarios of 6G networks are summarized as uHDD, uMUB and uHSLLC. The uMUB enables 6G networks to deliver any required performance within the space-aerial-terrestrial-underwater area, uHSLLC provides ultra-high rates and low latency, and uHDD meets the requirements of the data density and high-reliability. Similarly, in Ref. [10], the authors named the enhanced three scenarios in 6G networks as further-enhanced mobile broadband (FeMBB:), ultra-massive machine-type communications (umMTC) and enhanced-uRLLC (euRLLC). Apart from the three main scenarios, several other scenarios, such as extremely low-power communications (ELPC), long-distance and high-mobility communications (LDHMC) and intelligent IoT, are also promising in 6G networks$^{\rm [10], [33]}$.

\subsection{Core Requirements}

Emerging intelligent information services are the driving force behind the evolution of wireless communication networks. The rapid development of the emerging services, such as e-health, intelligent IoT and autonomous driving, results in an explosive growth in mobile data traffic, which will eventually exceed the limit of 5G networks. It is reported that 5G networks will reach its limits in a decade$^{\rm [7]}$. To meet the coming challenges, the KPIs for 6G networks will be$^{\rm [10]}$:
\begin{itemize}
\item
The peak data rate is at least $1$ Tb/s$^{\rm [34]}$, which is 50 times that of 5G networks. For THz-based 6G networks, the peak data rate is expected to reach up to $10$ Tb/s$^{\rm [34]}$.
\item
The user experience data rate is up to $1$ Gb/s, which is 10 times that of 5G networks. For some special scenarios, such as indoor hotspots, the user experience data rate is expected to reach up to $10$ Gb/s.
\item
The end-to-end latency is reduced to $10$-$100 \mu s$ and the mobility is at least $1000$ km/h, which is expected to provide potential applications for hyper-HSR and airline systems.
\item
The connection density is up to $10^7$ devices/$\rm km^2$ and the area traffic capacity is up to $1$ Gb/s$/\rm m^2$, which is 10 times that of 5G networks.
\item
The $10$-$100$ times energy efficiency and the $5$-$10$ times spectrum efficiency compared to 5G networks.
\end{itemize}

Upcoming intelligent information services have more stringent requirements for the KPIs of wireless communication networks. To satisfy typical scenarios and applications for the intelligent information society, 6G networks are required to provide superior network performance. Fig. \ref{Fig2} (b) compares the KPIs of 4G, 5G and 6G networks. In order to achieve the aforementioned system performance metrics, lots of novel enabling technologies, such as THz communications$^{\rm [34]}$, ultra-large-scale antenna arrays (i.e., UM-MIMO)$^{\rm [35]}$, OAM multiplexing$^{\rm [36], [37]}$ and IRS$^{\rm [38]}$ will be used in 6G networks. In many technologies of 6G networks, the most spectacular technology is IRS that can be widely deployed in 6G networks to achieve intelligent connectivity, ubiquitous connectivity, deep connectivity and holographic connectivity. In the following we will first introduce IRS technology and focus on the application of IRS in 6G networks.

\section{Intelligent Reflecting Surface: Theory and Core Advantages}

IRS is regarded as a promising technology that can intelligently reconfigure the channel environment with massive passive reflecting elements to effectively improve the performance of wireless communication networks. In this section, we summarize the IRS technology, involving the new IRS-assisted system model, the hardware architecture of IRS and competitive advantage in 6G networks.

\subsection{The IRS-assisted Two-ray System Model}

As multi-path propagation exists in a typical wireless communication environment, received signals are the compound signals with unpredictable time attenuation and propagation delay. Due to the constructive and destructive summation of compound signals, received signals will be significant distortions. This effect, terming as fading, is a major limitation in the next wireless communication systems. The main purpose of using IRS is to realize a reconfigurable wireless environment, in which the highly probabilistic wireless channel is  converted into a deterministic space by intelligently controlling the propagation of the electromagnetic (EM) waves in a software-controlled method$^{\rm [31]}$.

For easier analysis we first consider the two-ray channel model in the IRS-assisted single-input single-output (SISO) network as shown in Fig.\ref{Fig3}, in which the received signal consists of two components: the line-of-sight (LoS) ray and the ray reflected from IRS. In the proposed model, IRS is made of $N\times N$ reconfigurable metasurfaces each of which can independently control the amplitude and phase shift of reflection, the distance between the transmit and receive antennas is denoted as $d$, and the distances between the $(m,n)$-th element of IRS and the transmit and receive antennas are denoted as $r^{\text{T-I}}_{m,n}$ and $r^{\text{I-R}}_{m,n}$ respectively, $m,n=1,2,\cdots,N$. Without loss of generality, we assume unit gain transmit/receive antennas and a narrow-band transmission signal$^{\rm [31]}$, i.e., $x(t)=x(t-\tau_{m,n})$, where $x(t)$ is the complex baseband transmitted signal, $\tau_{m,n}=(r^{\text{T-I}}_{m,n}+r^{\text{I-R}}_{m,n}-d)/c$ is the relative time delay between the ray reflected from the $(m,n)$-th element of IRS and the LoS ray, $c$ is the speed of light. Then, the received baseband signal can be expressed as
\begin{align} \label{rt}
y(t)=&\frac{1}{2k}\bigg(\frac{e^{-ikd}}{d}+\sum_{m=1}^{N}\sum_{n=1}^{N}\frac{\beta_{m,n} e^{-ik(r^{\text{T-I}}_{m,n}+r^{\text{I-R}}_{m,n})}}{r^{\text{ T-I}}_{m,n}+r^{\text{I-R}}_{m,n}}\bigg)x(t) \nonumber\\
&+z(t),
\end{align}
where $i=\sqrt{-1}$ is the imaginary unit, $k=2\pi/\lambda$ is the wave number, $\lambda$ is the wavelength, $\beta_{m,n}=A_{m,n}e^{i\theta_{m,n}}$ is the programmable reflection coefficient of the $(m,n)$-th element in IRS, $A_{m,n}\in[0,1]$ and $\theta_{m,n}\in[0,2\pi]$ are the reconfigurable amplitude and phase of $\beta_{m,n}$ respectively, $m,n=1, 2, \cdots, N$, $z(t)$ is the related additive noise.

From \eqref{rt}, we can see that the received signal can be controlled and reconstructed spatially by IRS through smartly adjusting the reflection coefficients $\{\beta_{m,n}|m,n=1,2,\cdots,N\}$. For instance, each $\beta_{m,n}$ can be optimized so that the phase of the received signal from $N\times N$-element IRS is aligned with the phase of the LoS path, i.e., $\beta_{m,n}=A_{m,n}e^{i\theta_{m,n}}=e^{ik(r^{\text{T-I}}_{m,n}+r^{\text{I-R}}_{m,n}-d)}$, $m,n=1, 2, \cdots, N$, and then, the received signal with maximum power can be obtained whose power $P_r$ can be formulated as
\begin{align} \label{prmax}
P_r\overset{(a)}\approx P_t\big(N^2+1\big)^2\bigg(\frac{1}{4k^2d^2}\bigg),
\end{align}
where $P_t$ is the power of the transmitted signal $x(t)$, and (a) assumes the distance $d$ is large enough, i.e., $d\approx r^{\text{T-I}}_{m,n}+r^{\text{I-R}}_{m,n}$, $m,n=1, 2, \cdots, N$$^{\rm [31]}$.

In 6G networks, for further taking advantage of space division multiplexing, the number of antennas equipped on both the transmitter and receiver will be increased$^{\rm [5]}$. For example, the ultra-massive MIMO (UM-MIMO) technology with the transmit and receive antennas $(M_r, M_t) = (1024,1024)$ is utilized for THz communications in 6G networks$^{\rm [39]}$. For the LoS UM-MIMO network assisted by IRS, the received baseband signal vector can be written as
\begin{align} \label{rt6G}
\textbf{\emph{y}}(t) = (\textbf{\emph{H}}_r\textbf{\emph{B}}\textbf{\emph{H}}_t+\textbf{\emph{H}}_{\textrm{LoS}})\textbf{\emph{x}}(t) + \textbf{\emph{n}}(t)
\end{align}
where $\textbf{\emph{H}}_t$ is the $N\times M_t$ channel matrix from the transmitter to IRS, $\textbf{\emph{H}}_r$ is the $M_r\times N$ channel matrix from IRS to the receiver, $\textbf{\emph{H}}_{\rm LoS}$ is the $M_r\times M_t$ channel matrix from the transmitter to the receiver, $\textbf{\emph{x}}(t)$ and $\textbf{\emph{n}}(t)$ are the transmitted baseband signal vector and the corresponding noise vector, $\textbf{\emph{B}}=[\beta_{m,n}]_{N\times N}$ is the programmable reflection coefficient matrix of IRS. In the practical design of $\textbf{\emph{B}}$, many factors, such as elements' mutual coupling, noise and hardware imperfections, need to be considered, and their impact on the performance of IRS is still an ongoing research topic.

\begin{table*}[t]
\newcommand{\tabincell}[2]{\begin{tabular}{@{}#1@{}}#2\end{tabular}}
\small
\centering
\caption{Comparison of IRS with other related technologies.}
  \begin{tabular}{lcccccccc}
  \toprule
  \textbf{Technology} &\textbf{Role} &\textbf{Duplex} &\textbf{\tabincell{l}{ Power\\budget}} &\textbf{Noise} &\textbf{Compatibility} &\textbf{Interference} &\textbf{\tabincell{l}{\;Hardware\\Complexity}}  &\textbf{\tabincell{l}{\;\quad Energy\\Consumption}} \\
  \midrule
  \specialrule{0em}{3pt}{3pt}
  \textbf{IRS} &{Helper} &\tabincell{l}{\ \,Full\\Duplex} &\tabincell{l}{Passive\\\ \,Low} &No Noise &Very High &Very Low &Very Low &Low\\
  \specialrule{0em}{3pt}{3pt}
  \textbf{AF Relaying} &{Helper} &\tabincell{l}{Half/Full\\\ Duplex} &\tabincell{l}{Active\\\ High} &\tabincell{l}{Additive\\\ \;Noise} &Low &High & High & High \\
  \specialrule{0em}{3pt}{3pt}
  \textbf{DF Relaying} &{Helper} &\tabincell{l}{Half/Full\\\ Duplex} &\tabincell{l}{Active\\\ High} &\tabincell{l}{Additive\\\ \;Noise} & Low & High & High &High \\
  \specialrule{0em}{3pt}{3pt}
  \textbf{\tabincell{l}{Backscatter \\ Communications}} &{Source} &\tabincell{l}{\ \,Full\\Duplex} &\tabincell{l}{\ \ \,Active\\Very Low} &\tabincell{l}{Additive\\\ \;Noise} &Low &High &Low &Very Low\\
  \specialrule{0em}{3pt}{3pt}
  \bottomrule
  \label{Table1}
 \end{tabular}
\end{table*}

\subsection{Hardware Architecture}

The hardware implementation of IRS is based on the concept of ``metasurface", which is a kind of ultrathin man-made material with sub-wavelength elements. Specifically, the metasurface is a planar array comprising a mass of passive scattering elements whose EM properties depend on their structural parameters, such as the geometry (e.g., square, split-ring or hexagon), size, orientation and arrangement$^{\rm [3], [38]}$. It is reported in Ref. [40] that a planar metasurface  with $0.4 m^2$ and $1.5 mm$ thickness consists of $102$ controllable electromagnetic unit cells. Through proper external stimulation, the physical parameters of scattering elements can be altered, leading to the change of signal response of IRS without refabrication$^{\rm [41]}$.

The typical architecture of IRS is shown in Fig.\ref{Fig1}, which consists of an intelligent IRS controller and three layers$^{\rm [3]}$. In the outer layer, plenty of tunable elements embedded with metasurfaces are printed on the dielectric substrate to directly interact with incident signals. In wireless communication networks, the tunable elements are required to be able to reconfigure in real time to cater to dynamic channels arising from the user mobility$^{\rm [3]}$, which can be achieved by some electronic devices. One candidate is a positive-intrinsic negative (PIN) diode that can be switched between ``On" and ``Off" states by different direct-current (DC) voltage to generate $0$ and $\pi$ phase shifts$^{\rm [42]}$. Another candidate is a varactor diode that can be adjusted continuously by regulating the reverse bias voltage$^{\rm [43]}$. Moreover, to avoid the signal energy leakage, a copper backplane is set in the middle layer$^{\rm [3]}$. Last, the inner layer is a control network that is used to adjust the reflection coefficient of tunable elements. The control network can be an integrated network with multiple tiny controllers, in which each controller is only used to control one tunable element$^{\rm [44], [45]}$. Each tiny controller in the integrated network is able to interpret external instructions and adjust the configuration of the tunable element to achieve the EM reconfiguration of IRS. Moreover, the control network can also be made up of a smaller group of controller chips, in which each controller chip serves several tunable elements$^{\rm [46]}$. All of the controller chips are interconnected locally and communicate wirelessly to the IRS controller. In practice, a field programmable gate (FPGA)  can  be employed as not only the IRS controller but also a gateway to communicate and coordinate with other network devices, such as base stations(BSs) and terminals, through separate wireless links for low-rate information exchange with them$^{\rm [3], [38]}$.

\subsection{Technology Comparison and Performance Advantages}

When new technologies come into the spotlight, there is a responsibility to rigorously examine the potential benefits and limitations that they may offer compared to similar and mature technologies. Therefore, it is advisable to contrast IRS with transmission technologies that are likely to be bound up with them. The two technologies that are often deemed to be equivalent to IRS are relay-aided transmission and backscatter communications$^{\rm [31]}$. Hence, we will compare and analyze IRS, relay-aided communications and backscatter communications in the following.

The common relay-aided communications include amplify-and-forward (AF)  relay-aided communications and decode-and-forward (DF) relay-aided communications. Both relay-aided communication schemes require relay stations, which need a dedicated power source for operation, to process relay signals. The difference between AF relaying and DF relaying is that the relay signals are amplified and forwarded in the AF relay station$^{\rm [47]}$ but decoded and forwarded in the DF relay station$^{\rm [48]}$. However, both of signal processing have high complexity, thus leading to high hardware complexity and energy consumption$^{\rm [48], [49]}$. Moreover, the active electronic components used in relay stations are responsible for the presence of additive noise that negatively affects the performance of relaying communications$^{\rm [48]}$. Especially, the noise is amplified in AF relay stations.

Backscatter communications are a method for low-energy wireless communications using modulating signals scattered from a transponder (RF tag)$^{\rm [47]}$. Specifically, backscatter communications have no electronics for power harvesting or transmitting and only a switching mechanism or a variable load to modulate the wave reflections, which promises battery less or ultra-low power communications$^{\rm [50]}$. Besides, compared with the traditional schemes, the signal processing of backscatter communications is relatively simple, thus reducing the energy consumption and hardware complexity. However, the transmitted signal in backscatter communication systems is modulated and transmitted by the transponder, which may introduce additional noise.

It is worth noting that the proposed IRS differs significantly from all above techniques. Firstly, compared with relaying that assists transmission by actively generating new signals, IRS reflects the incident RF signals by the massive passive reflecting elements without the transmitter module$^{\rm [47]}$, which thus do not incurs additional power consumption and noise. For example, the IRS with 256 elements designed in Ref. [51] consumes only about $0.72$W, significantly lower than that of the active relaying in practice. Meanwhile, IRS usually operates in the full-duplex (FD) mode and thus has higher spectrally efficiency than relaying operating in the half-duplex (HD) mode. Secondly, different from traditional backscatter communications, IRS is utilized mainly to enhance the existing communication link performance instead of delivering its own information by reflection$^{\rm [47]}$. Therefore, the direct-path in IRS-assisted communications does not need to be suppressed and can also carry the same useful information as the reflect-path to maximize the total received power. Moreover, IRS possesses other advantages, such as easy deployment, good compatibility with existing networks and having full-band response$^{\rm [31], [52]}$. A more detailed comparison between the above technologies and IRS are summarized in Table \ref{Table1}.

Based on the above advantages, IRS naturally becomes a promising technology for the next 6G networks. In the following, we will mainly discuss the performance advantages of IRS-assisted wireless networks over traditional wireless networks without IRS to predict the performance of IRS-assisted 6G networks. In Ref. [53], an IRS-assisted THz MIMO system is proposed to mitigate blockage vulnerability in indoor scenarios, in which IRS is used to control the propagation direction of THz beam and enhance the coverage performance by intelligently adjusting the phase shifts of reflecting elements. The simulation results show that the data rate of IRS-assisted THz MIMO system is improved by about $45\%$ over the traditional THz system without IRS. In Ref. [54], a novel feedforward fully connected structure based deep neural network (DNN) scheme is designed and applied in IRS-assisted THz communications, which has the ability to output the optimal phase shift configurations assisted by the estimated channel parameters. The simulation results show that the DNN-based IRS scheme achieves a near-optimal communication rate performance, reaching $190\%$ of that in the non-IRS-assisted network. Besides, the authors in Ref. [55] design an IRS-assisted mobile edge computing (MEC) system that is capable of significantly outperforming the conventional MEC system operating without IRS. Quantitatively, about $20\%$ computational latency reduction is achieved over the conventional MEC system. Moreover, in Ref. [56], a multi-user multiple input single output (MISO) communication scheme with a large IRS is proposed, which can provide significant energy efficiency gains compared to conventional relay-assisted communications. Specifically, the energy efficiency of IRS-assisted system is about $45\%$ higher than that of the relay-assisted system. Meanwhile, the authors in Ref. [29] indicate that, compared with the conventional multi-antenna AF relaying, the IRS-assisted wireless network can provide up to $300\%$ energy efficiency. Moreover, the authors in Ref. [57] indicate deploying large-scale IRS in the point-to-point MISO communication system is more efficient than increasing the antenna array size for enhancing the spectral efficiency. Specifically, the average spectral efficiency of IRS-assisted system is improved by about $6.7\%$ compared with the traditional system without IRS. In Ref. [58], a new sum-path-gain maximization (SPGM) criterion is proposed and applied in the IRS-assisted point-to-point MIMO system, and the simulation results show that the spectral efficiency of IRS-assisted system with SPGM criterion is improved by nearly $40 \%$ compared with non-IRS-assisted system. In Ref. [59], the authors discuss the main performance gains of IRS-assisted massive MIMO non-orthogonal multiple access (NOMA) networks. Specifically, the achievable rate of the far user in IRS-NOMA system is improved by more than 6 times when compared with that achieved in MIMO-NOMA, and the energy efficiency of IRS-NOMA system is also greatly improved. Based on the above discussion, there are reasons to believe that compared with the existing 5G networks, the performance of IRS-assisted 6G networks, such as data rates, spectrum efficiency and energy efficiency, will be greatly improved due to the powerful environmental control capacity of IRS, and the main performance advantages of IRS-assisted networks are summarized in Table \ref{Table2}.

\begin{table}[t] 
\caption{Performance advantages of IRS-assisted wireless network} 
\begin{tabular}{p{2.8cm}p{5cm}} 
\hline
\hline
\textbf{Performance} & \textbf{IRS-assisted wireless network} \\ 
\hline 
Data rate & Up to $190\%$ of the non-IRS-assisted network$^{\rm [55]}$.\\
\hline
Latency & $20\%$ lower than the non-IRS-assisted network$^{\rm [55]}$.\\
\hline
Energy efficiency & Up to $300\%$ of the non-IRS-assisted network$^{\rm [29]}$.\\
\hline
Spectrum efficiency & $40\%$ higher than the non-IRS-assisted network$^{\rm [58]}$. \\
\hline
\hline
\label{Table2}
\end{tabular}
\setlength{\abovecaptionskip}{0cm}   
\setlength{\belowcaptionskip}{-1cm}   
\end{table}

\section{IRS-assisted 6G Wireless Networks}

With its flexibility in deployment and reconfiguration, low implementation cost, and low power consumption, IRS is expected to improve the transmission performance in the next 6G wireless networks. In this section, we propose a general idea of IRS-assisted 6G wireless networks including intelligent and controllable wireless environment, ubiquitous connectivity supported by IRS, IRS-assisted deep connectivity, and IRS-assisted holographic communications.

\subsection{Intelligent and Controllable Wireless Environment}
The first to the fifth generation wireless communication networks have been designed by assuming that wireless channels between communicating devices are decided by various uncontrollable physical environments, e.g., outdoor urban/suburban environment and indoor environment$^{\rm [60]-[62]}$. A number of channel models have been proposed for different environments and frequencies, which are assumed to cannot be modified and reconstructed, and can be only compensated through sophisticated transmission and reception schemes. However, in 6G wireless networks, the improvements that can be expected by operating only on the end-points of the wireless environment are not likely be sufficient to meet the challenging KPI requirements of ultra-high throughput, ultra-low latency, and ultra-high reliability. Therefore, IRS can be proposed as the breaking technology of turning the wireless environment into an optimization variable, which can be controlled and programmed rather than just adapted to.
\begin{figure}[tb]
  \centering
  \includegraphics[scale=0.32]{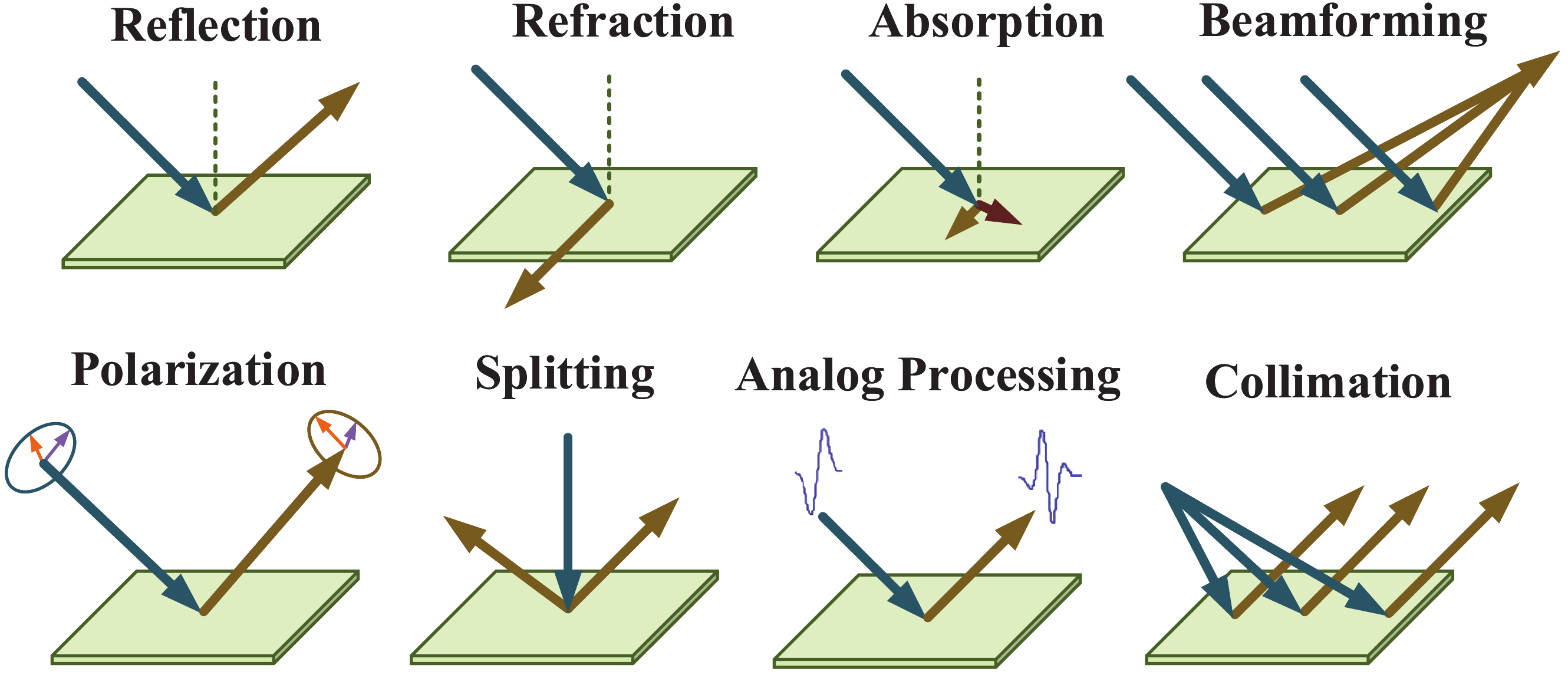}\\
  \caption{Electromagnetic-based elementary functions$^{\rm [63]}$.}
  \label{Fig5}
\end{figure}

The EM control functions of IRS include reflection, refraction, absorption, beamforming, polarization, splitting, analog processing, and collimation as illustrated in Fig.\ref{Fig5}. Reflection/refraction is reflecting/refracting incident radio waves to a specified direction that does not necessarily coincide with the incident direction$^{\rm [63]}$. Absorption means intelligently designing IRS that nulls, for the incident radio waves, the corresponding radio waves that are reflected and refracted. Beamforming is focusing incident radio waves toward a given location, i.e., concentrating the energy of incident radio waves. Polarization refers to modify the polarization modes of incident radio waves, for example, incident radio waves are transverse electric polarized, and reflected radio waves can be modified as transverse magnetic polarized$^{\rm [63]}$. Splitting means creating multiple reflected or refracted radio waves for the incident radio waves. Analog processing refers to directly perform mathematical operations at the EM level, e.g., the radio waves refracted by IRS can be the first-order derivative or the integral of incident radio waves$^{\rm [63]}$. Collimation can be regarded as the complementary of beamforming.

Moreover, IRS can also control the output signals through digital processing by the FPGA-based IRS controller. In this case, the IRS-assisted transmitter scheme can realize an IRS-based versions of spatial modulation (SM), index modulation (IM)$^{\rm [64]-[69]}$, and multi-antenna spatial multiplexing (MASM)$^{\rm [25], [26], [63], [70]-[72]}$. There is reason to believe that IRS-based MASM is attractive in 6G networks since multiple data streams can be transmitted simultaneously with a single feeder. Based on the above modes, several other IRS-assisted transmitter designs can be realized$^{\rm [63]}$, which can shape radio waves emitted by IRS through a simple RF feeder and an encoder.

\begin{figure}[tb]
  \centering
  \includegraphics[scale=0.40]{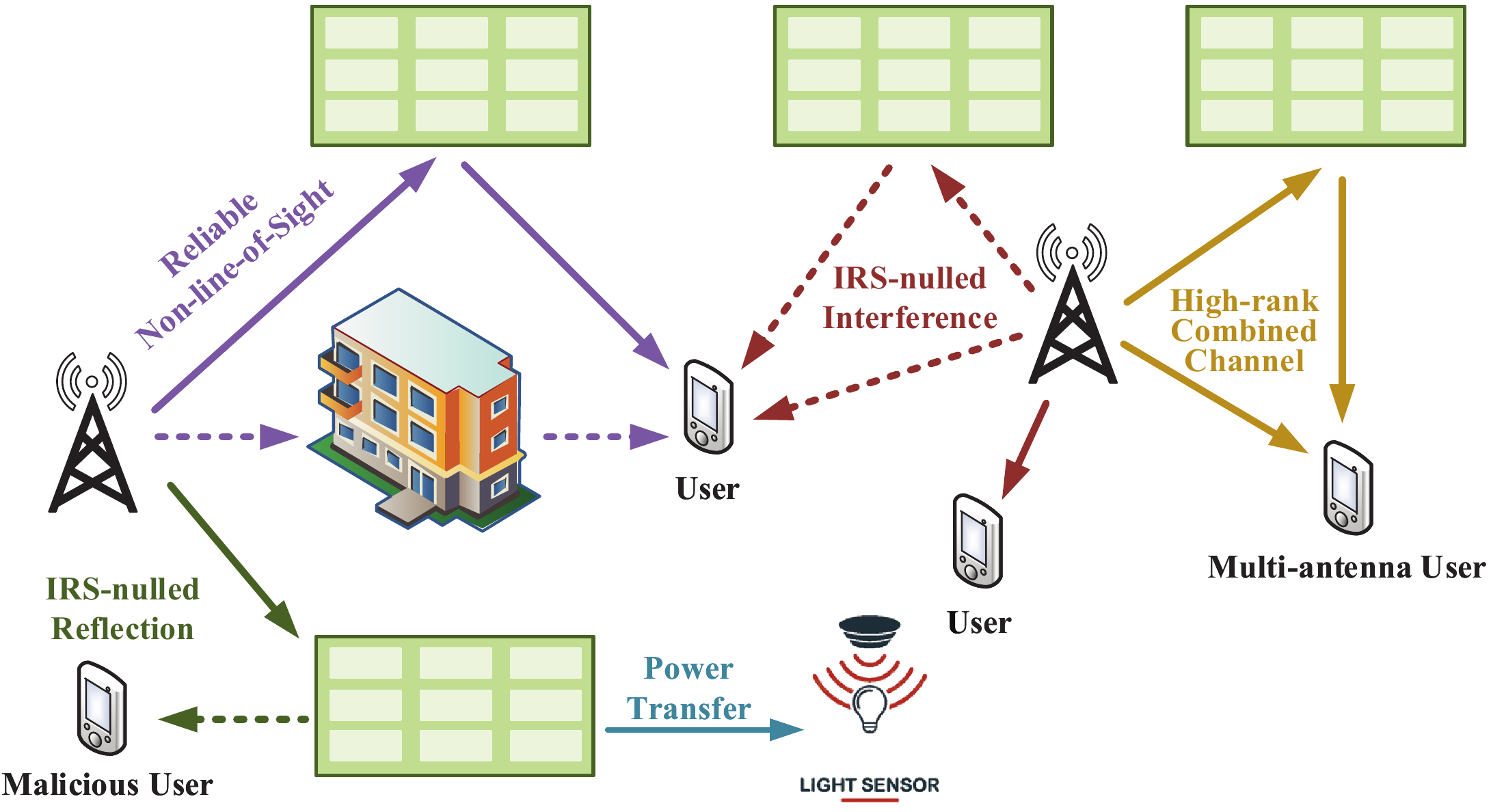}\\
  \caption{Potential applications of IRS in intelligent connectivity$^{\rm [63]}$.}
  \label{Fig10}
\end{figure}

Due to the EM and digital control capabilities of IRS, intelligent connectivity of 6G networks can be achieved assisted by IRS. Fig.\ref{Fig10} describes some potential applications of IRS in 6G networks. For example, IRS can be configured to generate adaptive non-line-of-sight (NLoS) links in dead zone or low coverage areas where line-of-sight (LoS) communications are impossible or insufficient due to obstructions/shadowing$^{\rm [73]-[75]}$. Meanwhile, IRS also can be configured to steer signals towards required directions or locations not only for enhancing the signal quality but also for suppressing unwanted signals that interfere with the wireless network$^{\rm [76]}$. Besides, IRS can be configured to worsen the signal towards malicious users either by creating destructive interference or by changing the reflection of signals off the locations occupied by malicious users$^{\rm [77]-[79]}$. Moreover, IRS can be intelligently configured to shape the wireless environment whose channel matrix has a high rank and a good condition number, to increase the channel capacity$^{\rm [80]}$. Furthermore, IRS can be configured to transmit wireless information and power to IoT devices at the same time. Other potential applications of IRS in 6G networks are summarized in Ref. [63].

Several potential scenarios in 6G networks can benefit from IRS-assisted intelligent connectivity. For example, in the city, a large mount of the outside wall of high buildings are made of glass. Smart glasses with special IRSs$^{\rm [63]}$ can be deployed to effectively control the propagation of radio waves. Besides, in crowded areas, such as large buildings, university campuses, offices, classrooms, IRS can be deployed to offer the desired high-speed connectivity without the cecessity of installing several access points. Most strikingly, the intelligent cloth can be realized by embedding IRSs and smart sensors to create wearable body area networks for monitoring the health of people in real time. These IRS-assisted applications set the foundation for the deployment of the intelligent information society.

\subsection{Ubiquitous Connectivity Supported by IRS}

With the rapid development of aerospace and exploration technologies, the scope of human activity is gradually expanding. In the next decade, more people will get a chance to get to space, and the communication requirements between satellites and ground/spacecraft will greatly increase. Meanwhile, the scope of human activity on the earth will be extended to many unfrequented areas including high-seas, desert hinterlands, and polar areas, more and more uninhabited islands will be settled by humans$^{\rm [14]}$. Therefore, the next 6G wireless networks are required to support ubiquitous connectivity to cover the space-aerial-terrestrial-underwater area.

Due to the relatively simple deployment and low energy consumption of IRS, IRS-assisted communications can be used to significantly enhance the coverage of 6G networks in several potential scenarios$^{\rm [63], [81]}$. For example, in the outdoor urban environment, the outside wall of buildings can be coated with IRS. This offers opportunities for enhancing the coverage and improving the energy efficiency of 6G networks in cities. Especially, IRS can be deployed upon the surface of high buildings, thereby creating a virtual LoS link between the access point and IRS,  which is particularly favorable when the direct path is severely shadowed$^{\rm [82]}$. Besides, due to the large deployment of IRS, the EM radiation caused by network infrastructures in the city could be greatly reduced. In the indoor environment, interior walls can also be coated with IRS for enhancing the local connectivity of several kinds of devices, such as mobile phones and tablets, which rely on wireless connectivity for operation. Especially, IRS can be deployed in ceilings and walls underground to provide the necessary connectivity to a large number of users simultaneously. Besides, the skin of cars/high-speed trains/airplanes can be coated with IRS, which can serve as moving nearly-passive relays for enhancing the coverage of 6G networks in the air and ground moving scenarios. Meanwhile, the interior and glasses of cars/high-speed trains/airplanes also can be coated with IRS that can provide high-speed Internet to passengers while reducing their EM field exposure and decreasing energy consumption. Moreover, IRS-assisted unmanned aerial vehicles (UAVs) communications$^{\rm [83]-[89]}$ can be employed in 6G networks to effectively enhance the coverage in remote areas, such as remote rural and desert areas. It is worth noting that IRS deployed in UAV networks can harvest energy from the incident signals, thus not requiring dedicated energy supply$^{\rm [87]}$. Furthermore, IRS can be employed in satellite communications$^{\rm [90]}$ and offshore communications$^{\rm [91]}$ to provide a cost-effective coverage of high-speed data services. There is reason to believe that by using the IRS, the coverage of 6G networks can be significantly enhanced, thus achieving the goal of covering the space-aerial-terrestrial-underwater area.

\subsection{IRS-Assisted Deep Connectivity}

Human production and living space are expanding continuously, and the types and scenarios of information interaction are becoming more and more complex. Internet of everything, starting with 5G networks, promotes rapid growth in IoT communication demand, which is likely to peak in the next decade. Compared with man-machine interaction, the information interaction in IoT will be greatly expanded in both information types and communication range. It can be predicted that in the future, the intelligent IoT will develop rapidly from the following two aspects: 1) the deep expansion of the activity space for connected devices, 2) the ability to deep interactive sensing, deep data mining and deep mind$^{\rm [14]}$. Accordingly, 6G networks are required to provide deep connectivity to support the upcoming intelligent IoT service.

To achieve the deep expansion of the activity space for intelligent IoT devices, the primary issue is how to continuously supply the energy to ubiquitous devices. Since IRS with the large aperture can provide continuous and stable EM energy to nearby devices by passive beamforming, IRS can be used in intelligent IoT networks to transmit wireless information and power to IoT devices simultaneously$^{\rm [3]}$. In Ref. [92], a semidefinite relaxation (SDR)-based algorithm is used in an IRS-assisted IoT network to maximize the weighted sum power harvested by devices. In Ref. [93], an IRS-assisted IoT network is proposed, whose efficiency of both downlink energy beamforming (EB) and uplink over-the-air computation (AirComp) is drastically enhanced by dynamically reconfiguring the propagation environment. Besides, the authors in Ref. [94] indicate the coverage of the IRS-assisted IoT network can be significantly improved due to the wireless energy transfer capability of IRS. These studies set the foundation for the application of IRS in the intelligent IoT network.

Moreover, due to powerful controlling ability, deep learning (DL) technologies$^{\rm [95]}$ are naturally applied in IRS to assist the IRS controller reconstructing the wireless environment, which also provides a basis for the application of IRS in intelligent IoT networks. The authors in Ref. [96] indicate both IRS and DL are the key to realize the intelligent wireless environment, and DL can be used in IRS-assisted networks to reduce the network complexity$^{\rm [96], [97]}$. In Ref. [98], a neural-network-based approach is used in the IRS controller to configure software-defined elements for creating intelligent wireless environments. In Ref. [99], an artificial neural network is used in the IRS-assisted indoor network to maximize the received power. Besides, a deep reinforcement learning framework (DRLF) is used in the IRS-assisted MISO wireless system to maximize the downlink received signal-to-noise ratio (SNR)$^{\rm [100]}$. In Ref. [101], a deep convolutional neural network is designed and trained in the IRS controller to restructure the programmable elements in milliseconds for intelligent multi-beam steering. In Ref. [102], a DRLF is used in the IRS-assisted MIMO system to achieve the joint design of transmit beamforming and element phase shifts in real time. In Ref. [103], a two-stage neural network is trained offline in the IRS-assisted MISO system in an unsupervised manner to maximize the system sum-rate with lower complexity. Furthermore, the authors in Ref. [104] propose the DRLF with reduced training overhead can be used in IRS controller to tune the phase shifts of IRS elements without the assistance of base stations or infrastructure nodes, which paves the way to the deployment of the distributed IRS in intelligent IoT networks. In Ref. [105], a fully connected artificial neural network is employed in the IRS-assisted wireless system to directly estimate the channels and phase angles from the reflected signals received by IRS, which enables the IRS-assisted wireless system to perform symbol detection without any dedicated pilot signaling significantly reducing the overhead. Additionally, compressive sensing and DL technologies are used in IRS with sparse channel sensors to reduce the overhead related to IRS control$^{\rm [106]}$. In Ref. [107], a twin convolutional neural network architecture is designed and used in massive MIMO systems with large IRS to estimate both direct and cascaded channels in the multi-terminal scenario. In Ref. [108], machine learning approaches are used in the IRS-assisted NOMA network to achieve IRS phase shift control, joint network deployment, and power allocation simultaneously for maximizing energy efficiency. Based on the above results, the IRS-assisted wireless network combined with DL technologies is expected to realize deep sensing and intelligent interaction with the wireless environment, and also has many other advantages including real-time beam reconstruction, high energy efficiency, and low overhead. These advantages lay the foundation for the application of DL-based IRS in intelligent IoT networks, making the controllable connection between intelligent devices possible. Therefore, the DL-based IRS can be considered to be a key enabler for realizing the vision of deep connectivity in 6G networks.

\subsection{IRS-Assisted Holographic Communications}
It is reported in Ref. [14] that in the next decade, the form of man-machine interaction is developed from the two-dimensional information interaction to high-fidelity VR/AR interaction and even holographic information interaction. VR/AR can create a virtual image in the real world by reconstructing the image of objects, that is, people can see both the virtual world and the real world at the same time. Therefore, 6G network are required to reconstruct wavefronts of waves in real time to support the high-fidelity VR/AR services without the limitation of location, and even holographic communication and imaging. It can be predicted that people can enjoy immersive holographic interaction services from anywhere at any time in the future, which is the vision of 6G networks ``holographic connectivity".

In theory, VR/AR or holographic display devices are required to reconstruct wavefronts of waves in real time to achieve a dynamic holographic display. However, the traditional holography has the issue of insufficient reconstruction ability$^{\rm [109]}$. Recently, many studies on metasurface have proposed a variety of metasurface holography that can completely control the amplitude and phase shift of waves in real time to solve the issue of traditional holography$^{\rm [110]-[112]}$. The above studies have laid the foundation for the application of IRS in holographic communication and imaging. Considering the ability to reconstruct the wavefront of waves in real time, IRS naturally becomes a promising technology for holographic communication and imaging in 6G networks. In Ref. [113], the IRS-based hologram is designed and realize for the first time, which can generate different holographic images with high-resolution and low-noise in real time by switching the phase shifts of recoding elements in IRS. Besides, the authors in Ref. [114] also propose and demonstrate an IRS-based hologram that can flexibly image by tunable phase elements of IRS. In Ref. [115], a high-efficiency holographic imaging method for IRS is proposed to achieve more powerful manipulations of EM waves, thus realizing real-time, continuous and flexible control of holographic images. In Ref. [116], a systematic approach to IRS hologram synthesis is presented, which exploits the rich field transformation capabilities of IRS for creating a variety of virtual images. In Ref. [117], the authors propose and analyze a real-time reconfigurable metasurface that can be used in IRS holography to realize dynamic holographic imaging at optical frequencies. Furthermore, an effective connection approach between IRS holograms and VR users is implemented in the VR network at THz frequency band$^{\rm [118]}$, which lays a foundation for the practical application of IRS holograms. Based on the above discussion, the IRS-based holography is key in enabling future interaction devices with reconfigurable image functionality, which can lead to advances in high-fidelity VR/AR services, and holographic communication and imaging.

\section{Challenges and Conclusions}

\subsection{New Challenges in IRS Application and Deployment}

In the implementation of IRS-assisted 6G networks, the first challenge is the design of tunable elements. Undoubtedly, continuously adjusting the reflection coefficients of each element is beneficial for the network performance, however, it is very costly to implement due to the sophisticated design and expensive hardware of massive high-precision elements. Therefore, a more cost-effective solution for IRS-assisted networks is to use tunable elements with discrete amplitude/phase shift levels, for instance, each element can only adjust two-level amplitude (reflecting or absorbing) and phase-shift ($0$ or $\pi$)$^{\rm [119]}$. Nevertheless, whether IRS with two-level control can meet the requirements of 6G networks needs further verification. Another challenge for IRS-assisted 6G networks is developing an effective control mechanism to connect and communicate with massive tunable elements, and thus agilely and jointly control their EM behaviors on demand. To date, although several control mechanisms have been proposed$^{\rm [120]-[123]}$, the control mechanism of IRS with ultra-large-scale or ultra-dense tunable elements is still an open issue in 6G networks. Generally, the IRS without any radio frequency (RF) chains is not able to perform any baseband processing functionality. Therefore, the channel between the transmitter and IRS  and the channel between IRS and receiver cannot be separately estimated through traditional training-based approaches. Recently, there have been several works taking various approaches to efficiently estimate the channels for IRS-assisted communication$^{\rm [3], [77], [78], [124]-[127]}$. However, in terms of the IRS reflected channels,  its sparsity and other properties in practical environments are still unknown. A channel estimation framework for a more general channel model without assuming any channel property is expected to be proposed. At the same time, since the transmit power and reflection coefficients are highly coupled, the  joint optimization of the channel assignment, power allocation and reflection coefficients for IRS systems is usually quite difficult. Hence, the efficient resource allocation is non-trivial to solve for IRS systems$^{\rm [128]}$.

So far, the research on the ubiquitous connectivity supported by IRS is still in the early theoretical stage, and thus more proof-of-concept prototypes are required to validate the IRS's practical efficiency. Moreover, there are a number of key issues in the development of IRS-assisted 6G holographic communications that need to
be resolved, such as IRS-assisted 3D holographic imaging, the EM reconstruction mechanism of IRS-based holograms in different 6G scenarios and the deployment of IRS-based holograms in 6G networks. In addition, the practical deployment of IRS on buildings in the city will involve different units and property managers, as well as the division of interests between operators and equipment providers, which also brings new challenges to the commercialization of IRS.

\subsection{Conclusions}

The intelligent information society with highly digitized, intelligence inspired and globally data driven will be deployed in the next decade, which creates core requirements for 6G networks, that is, intelligent connectivity, ubiquitous connectivity, deep connectivity, and holographic connectivity. Due to its powerful wireless environmental control capability, IRS becomes a promising technology for 6G networks. Specifically, IRS is capable of sensing the wireless environment and modulating its reflection coefficients dynamically to realize various functions. In this paper, we present a comprehensive survey on the theory and applications of IRS to the next 6G wireless communication networks. There is reason to believe that integrating IRS into 6G networks will fundamentally change the wireless network paradigm from an uncontrollable wireless environment to an intelligent and controllable wireless environment, thus opening fertile directions for future research. Since IRS-assisted wireless networks are new and remain largely unexplored, it is hoped that this paper could provide a useful and effective guide for future research on them.

\section*{Acknowledgment}
The authors would like to thank the editor and the anonymous reviewers for their careful reading and valuable suggestions that helped to improve the quality of this manuscript.


\begin{IEEEbiography}[{\includegraphics[width=1in,height=1.25in,clip,keepaspectratio]{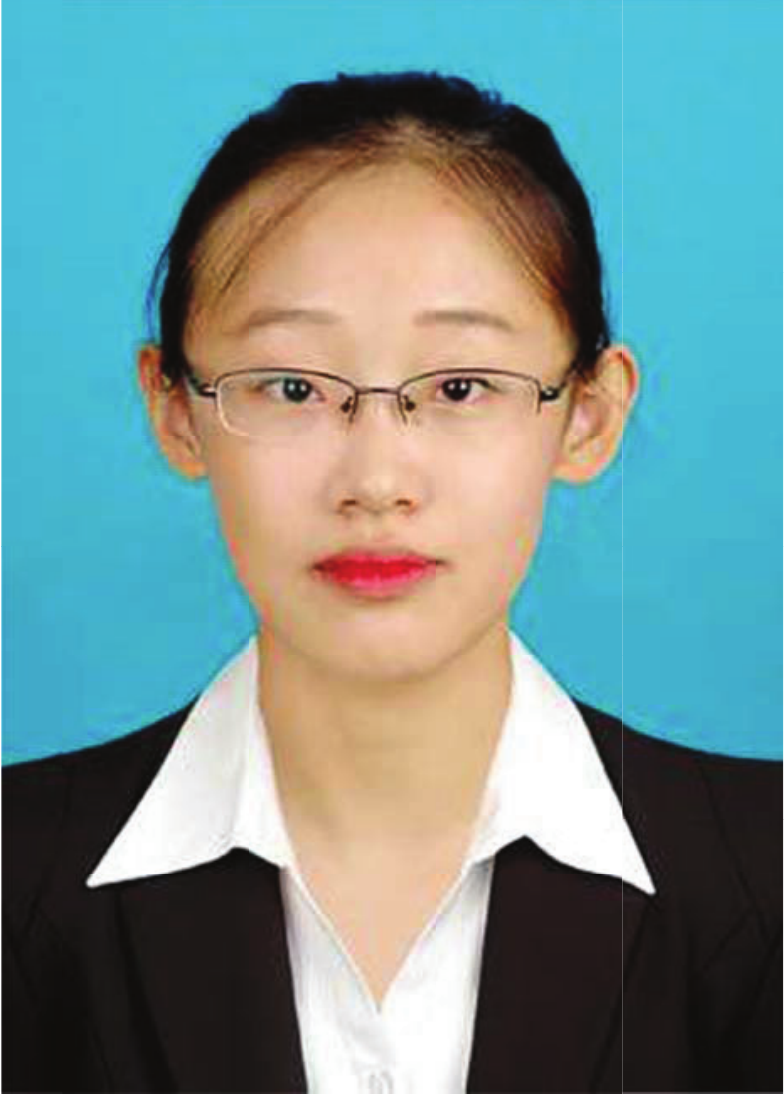}}]{Wen-Xuan Long}
(S'18) received the B.S. degree (with Highest Hons.) in Rail Transit Signal and Control from Dalian Jiaotong University, Dalian, China in 2017. She is currently pursuing a double Ph.D. degree in Communications and Information Systems at Xidian University, China and University of Pisa, Italy. Her research interests include broadband wireless communication systems and array signal processing.
\end{IEEEbiography}

\begin{IEEEbiography}[{\includegraphics[width=1in,height=1.25in,clip,keepaspectratio]{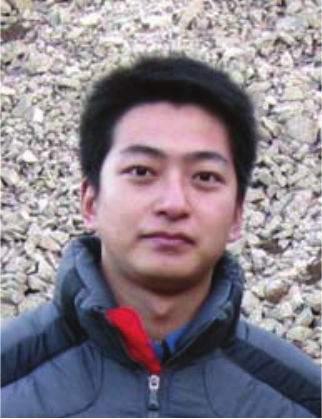}}]{Rui Chen}[corresponding author]
(S'08-M'11) received the B.S., M.S. and Ph.D. degrees in Communications and Information Systems from Xidian University, Xi'an, China, in 2005, 2007 and 2011, respectively. From 2014 to 2015, he was a visiting scholar at Columbia University in the City of New York. He is currently an associate professor and Ph.D. supervisor in the school of Telecommunications Engineering at Xidian University. He has published about 50 papers in international journals and conferences and held 10 patents. He is an Associate Editor for International Journal of Electronics, Communications, and Measurement Engineering (IGI Global). His research interests include broadband wireless communication systems, array signal processing and intelligent transportation systems.
\end{IEEEbiography}

\begin{IEEEbiography}[{\includegraphics[width=1in,height=1.25in,clip,keepaspectratio]{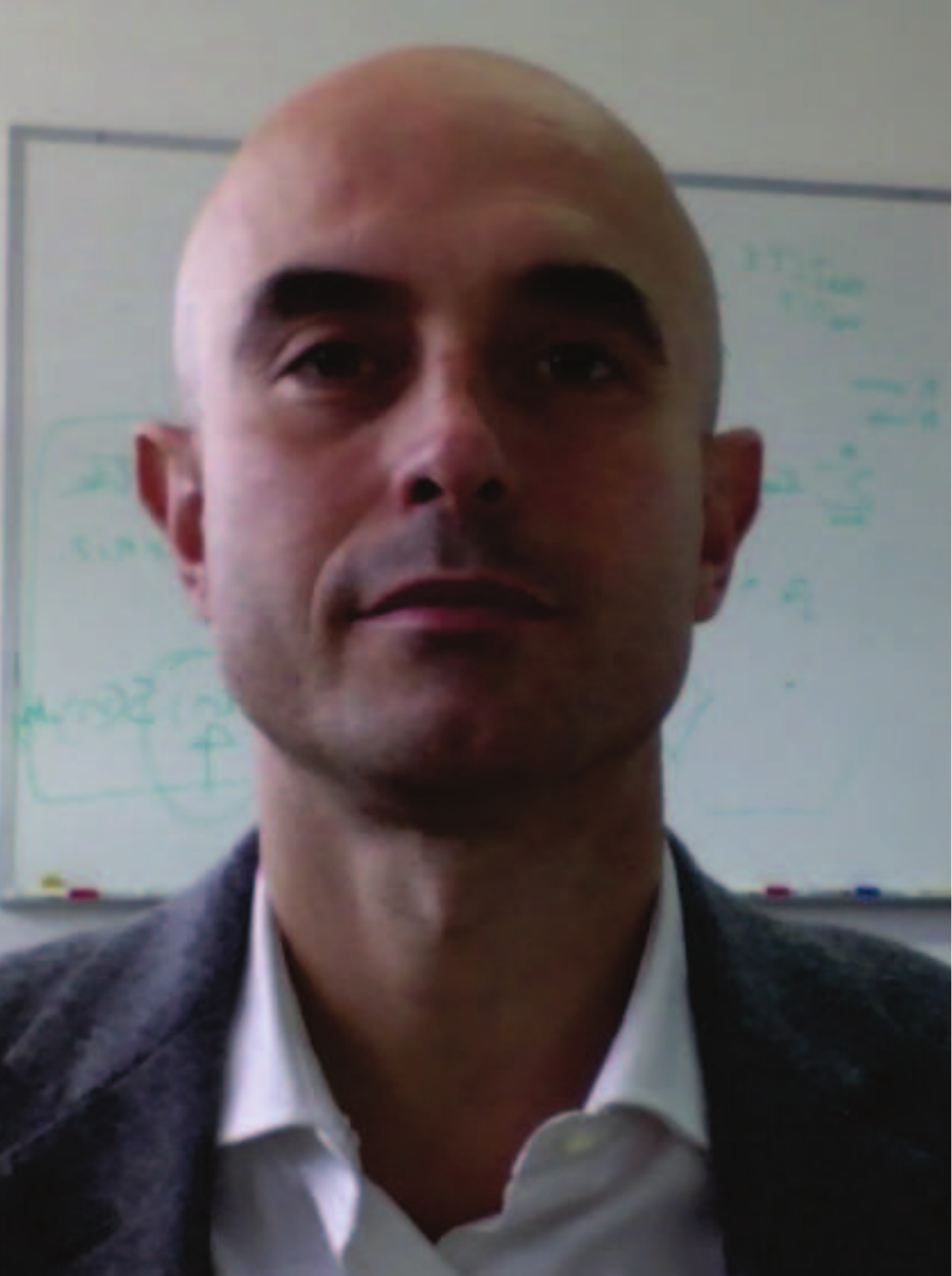}}]{Marco Moretti}
(M'03) received the degree in electronic engineering from the University of Florence, Florence, Italy, in 1995, and the Ph.D. degree from the Delft University of Technology, Delft, The Netherlands, in 2000. From 2000 to 2003, he was a Senior Researcher with Marconi Mobile. He is currently an Associate Professor with the University of Pisa, Pisa, Italy. His research interests include resource allocation for multicarrier systems, synchronization, and channel estimation. He is currently an Associate Editor of the IEEE TRANSACTIONS ON SIGNAL PROCESSING.
\end{IEEEbiography}

\begin{IEEEbiography}[{\includegraphics[width=1in,height=1.25in,clip,keepaspectratio]{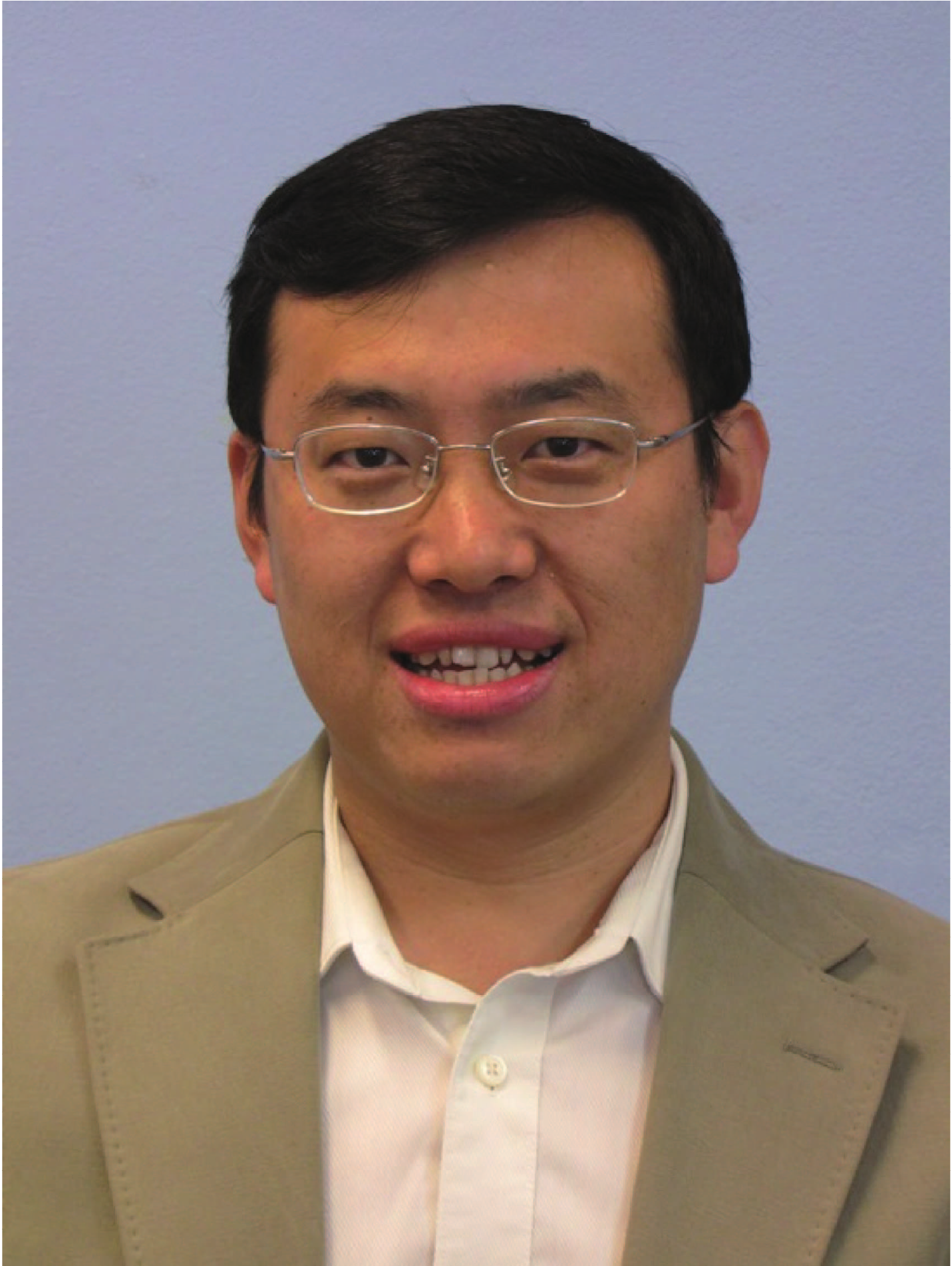}}]{Wei Zhang}
(S'01-M'06-SM'11-F'15) received the Ph.D. degree in Electronic Engineering from the Chinese University of Hong Kong in 2005. Currently, he is Professor at School of Electrical Engineering and Telecommunications, the University of New South Wales, Sydney, Australia. His research interests include space information networks, IoT, and massive MIMO. He has published over 200 papers and holds 5 US patents. He received 5 best paper awards from international conferences, IEEE Communications Society (ComSoc) TCCN Publication Award, and ComSoc Asia Pacific Outstanding Paper Award.

He was ComSoc Distinguished Lecturer in 2016-2017 and the Editor-in-Chief of IEEE Wireless Communications Letters in 2016-2019. Currently, he serves as Area Editor of IEEE Transactions on Wireless Communications and Editor-in-Chief of Journal of Communications and Information Networks. He also serves in Steering Committee of IEEE Networking Letters and IEEE Transactions on Green Communications and Networking.  He is Chair for IEEE Wireless Communications Technical Committee and Vice Director of the IEEE ComSoc Asia Pacific Board. He was a member of Fellow Evaluation Committee of IEEE Vehicular Technology Society in 2016-2019. He served on the organizing committee of the IEEE ICASSP 2016 and the IEEE GLOBECOM 2017. He was TPC Chair of APCC 2017 and ICCC 2019. He is a member of ComSoc IT Committee, TC Recertification Committee, and Finance Standing Committee. He is Member-at-Large of ComSoc in 2018-2020.
\end{IEEEbiography}

\begin{IEEEbiography}[{\includegraphics[width=1in,height=1.25in,clip,keepaspectratio]{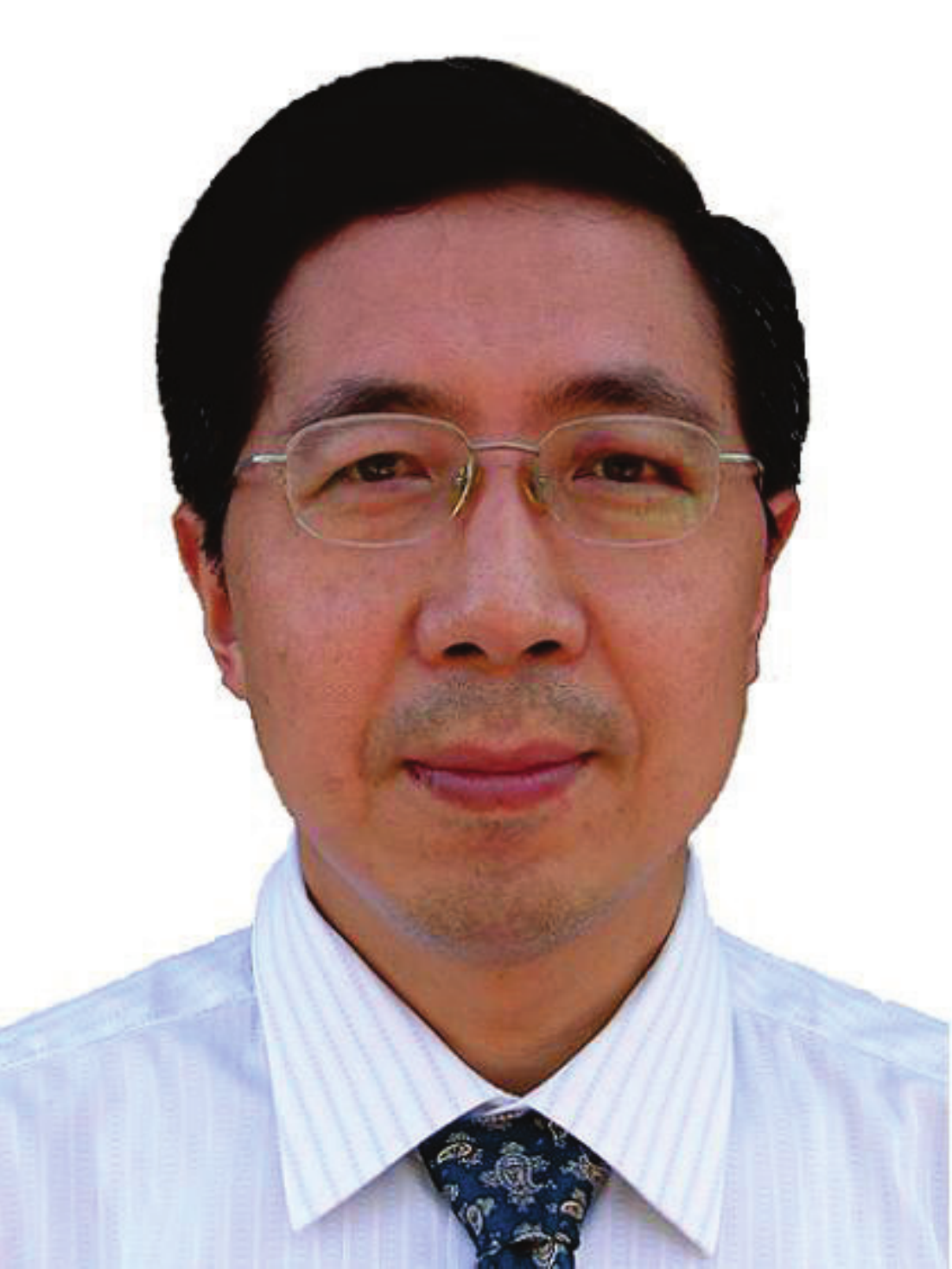}}]{Jiandong Li}
(SM'05) received the B.E., M.S., and Ph.D. degrees in communications engineering from
Xidian University, Xi'an, China, in 1982, 1985, and 1991, respectively. He was a Visiting Professor with the Department of Electrical and Computer Engineering, Cornell University, from 2002 to 2003. He has been a faculty member of the School of Telecommunications Engineering, Xidian University, since 1985, where he is currently a Professor and the Vice Director of the Academic Committee, State Key Laboratory of Integrated Service Networks. His major research interests include wireless communication theory, cognitive radio, and signal processing. He was recognized as a Distinguished Young Researcher by NSFC and a Changjiang Scholar by the Ministry of Education, China, respectively. He served as the General Vice Chair of ChinaCom 2009 and the TPC Chair of the IEEE ICCC 2013.
\end{IEEEbiography}

\begin{thebibliography}{200}
\bibitem{ref1}
Shafi M , Molisch A F , Smith P J , et al. 5G: A Tutorial Overview of Standards, Trials, Challenges, Deployment and Practice[J]. IEEE Journal on Selected Areas in Communications, 2017, 35(6):1201-1221.

\bibitem{ref2}
Wang C X, Haider F, Gao X, et al. Cellular architecture and key technologies for 5G wireless communication networks[J]. IEEE communications magazine, 2014, 52(2): 122-130.

\bibitem{ref3}
Wu Q, Zhang R. Towards smart and reconfigurable environment: Intelligent reflecting surface aided wireless network[J]. IEEE Communications Magazine, 2019, 58(1): 106-112.

\bibitem{ref4}
Wan Z, Gao  Z, M. Alouini. Broadband Channel Estimation for Intelligent Reflecting Surface Aided mmWave Massive MIMO Systems[C]//Proceedings of IEEE International Conference on Communications. Piscataway: IEEE Press, 2020:1-6.

\bibitem{ref5}
Yang P, Xiao Y, Xiao M, et al. 6G wireless communications: Vision and potential techniques[J]. IEEE Network, 2019, 33(4): 70-75.

\bibitem{ref6}
Katz M, Matinmikko-Blue M, Latva-Aho M. 6Genesis flagship program: Building the bridges towards 6G-enabled wireless smart society and ecosystem[C]//Proceedings of the IEEE 10th Latin-American Conference on Communications. Piscataway: IEEE Press, 2018: 1-9.

\bibitem{ref7}
Huang T, Yang W, Wu J, et al. A survey on green 6G network: Architecture and technologies[J]. IEEE Access, 2019, 7: 175758-175768.

\bibitem{ref8}
Latva-aho M. Radio access networking challenges towards 2030[J]. Powerpoint Presentation, 2018.

\bibitem{ref9}
Zong B, Fan C, Wang X, et al. 6G technologies: Key drivers, core requirements, system architectures, and enabling technologies[J]. IEEE Vehicular Technology Magazine, 2019, 14(3): 18-27.

\bibitem{ref10}
Zhang Z, Xiao Y, Ma Z, et al. 6G wireless networks: Vision, requirements, architecture, and key technologies[J]. IEEE Vehicular Technology Magazine, 2019, 14(3): 28-41.

\bibitem{ref11}
Tse D, Viswanath P. Fundamentals of wireless communication[M]. Cambridge university press, 2005.

\bibitem{ref12}
Rappaport T S. Wireless communications: principles and practice[M]. New Jersey: prentice hall PTR, 1996.

\bibitem{ref13}
You X, Wang C X, Huang J, et al. Towards 6G wireless communication networks: Vision, enabling technologies, and new paradigm shifts[J]. Science China Information Sciences, 2021, 64(1): 1-74.

\bibitem{ref14}
Zhao Y, Yu G, Xu H. 6G mobile communication network: vision, challenges and key technologies[J]. arXiv preprint arXiv:1905.04983, 2019.

\bibitem{ref15}
Yuan Y, Zhao Y, Zong B, et al. Potential key technologies for 6G mobile communications[J]. Science China Information Sciences, 2020, 63(8): 1-19.

\bibitem{ref16}
Tomkos I, Klonidis D, Pikasis E, et al. Toward the 6G network era: Opportunities and challenges[J]. IT Professional, 2020, 22(1): 34-38.

\bibitem{ref17}
Viswanathan H, Mogensen P E. Communications in the 6G era[J]. IEEE Access, 2020, 8: 57063-57074.

\bibitem{ref18}
Wang P, Fang J, Yuan X, et al. Intelligent reflecting surface-assisted millimeter wave communications: Joint active and passive precoding design[J]. IEEE Transactions on Vehicular Technology, 2020.

\bibitem{ref19}
Ning B, Chen Z, Chen W, et al. Terahertz multi-user massive MIMO with intelligent reflecting surface: Beam training and hybrid beamforming[J]. arXiv preprint arXiv:1912.11662, 2019.

\bibitem{ref20}
Hu S, Rusek F, Edfors O. The potential of using large antenna arrays on intelligent surfaces[C]//Proceedings of the IEEE 85th Vehicular Technology Conference. Piscataway: IEEE Press, 2017: 1-6.

\bibitem{ref21}
Hu S, Rusek F, Edfors O. Beyond massive MIMO: The potential of positioning with large intelligent surfaces[J]. IEEE Transactions on Signal Processing, 2018, 66(7): 1761-1774.

\bibitem{ref22}
Hu S, Rusek F, Edfors O. Beyond massive MIMO: The potential of data transmission with large intelligent surfaces[J]. IEEE Transactions on Signal Processing, 2018, 66(10): 2746-2758.

\bibitem{ref23}
Liaskos C, Nie S, Tsioliaridou A, et al. A new wireless communication paradigm through software-controlled metasurfaces[J]. IEEE Communications Magazine, 2018, 56(9): 162-169.

\bibitem{ref24}
Liaskos C, Nie S, Tsioliaridou A, et al. Realizing wireless communication through software-defined hypersurface environments[C]//Proceedings of the IEEE 19th International Symposium on ``A World of Wireless, Mobile and Multimedia Networks". Piscataway: IEEE Press, 2018: 14-15.

\bibitem{ref25}
Tang W, Li X, Dai J Y, et al. Wireless communications with programmable metasurface: Transceiver design and experimental results[J]. China Communications, 2019, 16(5): 46-61.

\bibitem{ref26}
Tang W, Dai J Y, Chen M, et al. Programmable metasurface-based RF chain-free 8PSK wireless transmitter[J]. Electronics Letters, 2019, 55(7): 417-420.

\bibitem{ref27}
Huang C, Zappone A, Alexandropoulos G C, et al. Reconfigurable intelligent surfaces for energy efficiency in wireless communication[J]. IEEE Transactions on Wireless Communications, 2019, 18(8): 4157-4170.

\bibitem{ref28}
Basar E. Reconfigurable intelligent surface-based index modulation: A new beyond MIMO paradigm for 6G[J]. IEEE Transactions on Communications, 2020, 68(5): 3187-3196.

\bibitem{ref29}
Dai L, Wang B, Wang M, et al. Reconfigurable intelligent surface-based wireless communications: Antenna design, prototyping, and experimental results[J]. IEEE Access, 2020, 8: 45913-45923.

\bibitem{ref30}
Hu J, Zhang H, Di B, et al. Reconfigurable intelligent surface based RF sensing: Design, optimization, and implementation[J]. IEEE Journal on Selected Areas in Communications, 2020, 38(11): 2700-2716.

\bibitem{ref31}
Basar E, Di Renzo M, De Rosny J, et al. Wireless communications through reconfigurable intelligent surfaces[J]. IEEE access, 2019, 7: 116753-116773.

\bibitem{ref32}
Yastrebova A, Kirichek R, Koucheryavy Y, et al. Future networks 2030: Architecture \& requirements[C]//Proceedings of the International Congress on Ultra Modern Telecommunications and Control Systems and Workshops. Piscataway: IEEE Press, 2018: 1-8.

\bibitem{ref33}
Zhang L, Liang Y C, Niyato D. 6G Visions: Mobile ultra-broadband, super internet-of-things, and artificial intelligence[J]. China Communications, 2019, 16(8): 1-14.

\bibitem{ref34}
Boulogeorgos A A A, Alexiou A, Merkle T, et al. Terahertz technologies to deliver optical network quality of experience in wireless systems beyond 5G[J]. IEEE Communications Magazine, 2018, 56(6): 144-151.

\bibitem{ref35}
Bj$\ddot{\rm o}$rnson E, Sanguinetti L, Wymeersch H, et al. Massive MIMO is a reality-What is next? Five promising research directions for antenna arrays[J]. Digital Signal Processing, 2019, 94: 3-20.

\bibitem{ref36}
Chen R, Zhou H, Moretti M, et al. Orbital angular momentum waves: generation, detection, and emerging applications[J]. IEEE Communications Surveys \& Tutorials, 2019, 22(2): 840-868.

\bibitem{ref37}
Chen R, Long W X, Wang X, et al. Multi-Mode OAM Radio Waves: Generation, Angle of Arrival Estimation and Reception With UCAs[J]. IEEE Transactions on Wireless Communications, 2020, 19(10): 6932-6947.

\bibitem{ref38}
Gong S, Lu X, Hoang D T, et al. Toward Smart Wireless Communications via Intelligent Reflecting Surfaces: A Contemporary Survey[J]. IEEE Communications Surveys \& Tutorials, 2020, 22(4): 2283-2314.

\bibitem{ref39}
Akyildiz I F, Han C, Nie S. Combating the distance problem in the millimeter wave and terahertz frequency bands[J]. IEEE Communications Magazine, 2018, 56(6): 102-108.

\bibitem{ref40}
Kaina N, Dupre M, Lerosey G, et al. Shaping complex microwave fields in reverberating media with binary tunable metasurfaces[J]. Scientific reports, 2014, 4(1): 1-8.

\bibitem{ref41}
Liu F, Pitilakis A, Mirmoosa M S, et al. Programmable metasurfaces: State of the art and prospects[C]//Proceedings of the IEEE International Symposium on Circuits and Systems. Piscataway: IEEE Press, 2018: 1-5.

\bibitem{ref42}
Cui T J, Qi M Q, Wan X, et al. Coding metamaterials, digital metamaterials and programmable metamaterials[J]. Light: Science \& Applications, 2014, 3(10): e218-e218.

\bibitem{ref43}
Zhao J, Cheng Q, Chen J, et al. A tunable metamaterial absorber using varactor diodes[J]. New Journal of Physics, 2013, 15(4): 043049.

\bibitem{ref44}
Liaskos C, Tsioliaridou A, Pitsillides A, et al. Design and development of software defined metamaterials for nanonetworks[J]. IEEE Circuits and Systems Magazine, 2015, 15(4): 12-25.

\bibitem{ref45}
Abadal S, Liaskos C, Tsioliaridou A, et al. Computing and communications for the software-defined metamaterial paradigm: A context analysis[J]. IEEE access, 2017, 5: 6225-6235.

\bibitem{ref46}
Tasolamprou A C, Mirmoosa M S, Tsilipakos O, et al. Intercell wireless communication in software-defined metasurfaces[C]//Proceedings of the IEEE International Symposium on Circuits and Systems. Piscataway: IEEE Press, 2018: 1-5.

\bibitem{ref47}
Wu Q, Zhang R. Intelligent reflecting surface enhanced wireless network via joint active and passive beamforming[J]. IEEE Transactions on Wireless Communications, 2019, 18(11): 5394-5409.

\bibitem{ref48}
Di Renzo M, Ntontin K, Song J, et al. Reconfigurable intelligent surfaces vs. relaying: Differences, similarities, and performance comparison[J]. IEEE Open Journal of the Communications Society, 2020, 1: 798-807.

\bibitem{ref49}
Bj$\ddot{\rm o}$rnson E, $\ddot{\rm O}$zdogan $\ddot{\rm O}$, Larsson E G. Intelligent reflecting surface versus decode-and-forward: How large surfaces are needed to beat relaying?[J]. IEEE Wireless Communications Letters, 2019, 9(2): 244-248.

\bibitem{ref50}
Khaleghi A, Hasanvand A O, Balasingham I. Wireless backscatter communication using multiple transmitter scheme[C]//Proceedings of the 10th European Conference on Antennas and Propagation. Piscataway: IEEE Press, 2016: 1-4.

\bibitem{ref51}
Tang W, Chen M Z, Chen X, et al. Wireless communications with reconfigurable intelligent surface: Path loss modeling and experimental measurement[J]. IEEE Transactions on Wireless Communications, 2020.

\bibitem{ref52}
Hu S, Rusek F, Edfors O. Beyond massive MIMO: The potential of data transmission with large intelligent surfaces[J]. IEEE Transactions on Signal Processing, 2018, 66(10): 2746-2758.

\bibitem{ref53}
Chen Z, Chen W, Ma X, et al. Taylor Expansion Aided Gradient Descent Schemes for IRS-Enabled Terahertz MIMO Systems[C]//Proceedings of the IEEE Wireless Communications and Networking Conference Workshops. Piscataway: IEEE Press, 2020: 1-7.

\bibitem{ref54}
Ma X, Chen Z, Chen W, et al. Joint channel estimation and data rate maximization for intelligent reflecting surface assisted terahertz MIMO communication systems[J]. IEEE Access, 2020, 8: 99565-99581.

\bibitem{ref55}
Bai T, Pan C, Deng Y, et al. Latency minimization for intelligent reflecting surface aided mobile edge computing[J]. IEEE Journal on Selected Areas in Communications, 2020, 38(11): 2666-2682.

\bibitem{ref56}
Huang C, Alexandropoulos G C, Zappone A, et al. Energy efficient multi-user MISO communication using low resolution large intelligent surfaces[C]//Proceedings of the IEEE Globecom Workshops. Piscataway: IEEE Press, 2018: 1-6.

\bibitem{ref57}
Yu X, Xu D, Schober R. MISO wireless communication systems via intelligent reflecting surfaces[C]//Proceedings of the IEEE/CIC International Conference on Communications in China. Piscataway: IEEE Press, 2019: 735-740.

\bibitem{ref58}
Ning B, Chen Z, Chen W, et al. Beamforming optimization for intelligent reflecting surface assisted MIMO: A sum-path-gain maximization approach[J]. IEEE Wireless Communications Letters, 2020, 9(7): 1105-1109.

\bibitem{ref59}
de Sena A S, Carrillo D, Fang F, et al. What role do intelligent reflecting surfaces play in multi-antenna non-orthogonal multiple access?[J]. IEEE Wireless Communications, 2020, 27(5): 24-31.

\bibitem{ref60}
Ghosh M, Srinivasarao C, Sahoo H K. Adaptive channel estimation in MIMO-OFDM for indoor and outdoor environments[C]//International Conference on Wireless Communications, Signal Processing and Networking. Piscataway: IEEE Press, 2017: 2743-2747.

\bibitem{ref61}
Hemrungrote S, Hori T, Fujimoto M, et al. Effects of path visibility on channel capacity of urban MIMO systems[C]//IEEE Radio and Wireless Symposium. Piscataway: IEEE Press, 2010: 653-656.

\bibitem{ref62}
de Britto Freire A P B, dos Santos Cavalcante G P. BER Estimation for DS-Spread Spectrum Systems in Densely Foliaged Suburban Areas[J] IEEE Ninth International Symposium on Spread Spectrum Techniques and Applications, 2006: 347-350.

\bibitem{ref63}
Di Renzo M, Zappone A, Debbah M, et al. Smart radio environments empowered by reconfigurable intelligent surfaces: How it works, state of research, and the road ahead[J]. IEEE Journal on Selected Areas in Communications, 2020, 38(11): 2450-2525.

\bibitem{ref64}
Di Renzo M, Haas H, Grant P M. Spatial modulation for multiple-antenna wireless systems: A survey[J]. IEEE Communications Magazine, 2011, 49(12): 182-191.

\bibitem{ref65}
Di Renzo M, Haas H, Ghrayeb A, et al. Spatial modulation for generalized MIMO: Challenges, opportunities, and implementation[J]. Proceedings of the IEEE, 2013, 102(1): 56-103.

\bibitem{ref66}
Yang P, Xiao Y, Guan Y L, et al. Single-carrier SM-MIMO: A promising design for broadband large-scale antenna systems[J]. IEEE Communications Surveys \& Tutorials, 2016, 18(3): 1687-1716.

\bibitem{ref67}
Basar E, Wen M, Mesleh R, et al. Index modulation techniques for next-generation wireless networks[J]. IEEE access, 2017, 5: 16693-16746.

\bibitem{ref68}
Di Renzo M. Spatial modulation based on reconfigurable antennas-A new air interface for the IoT[C]//IEEE Military Communications Conference. Piscataway: IEEE Press, 2017: 495-500.

\bibitem{ref69}
Wen M, Zheng B, Kim K J, et al. A survey on spatial modulation in emerging wireless systems: Research progresses and applications[J]. IEEE Journal on Selected Areas in Communications, 2019, 37(9): 1949-1972.

\bibitem{ref70}
Tang W, Dai J Y, Chen M, et al. Subject Editor spotlight on programmable metasurfaces: The future of wireless?[J]. Electron. Lett., 2019, 55(7): 360-361.

\bibitem{ref71}
Tang W, Chen M Z, Dai J Y, et al. Wireless communications with programmable metasurface: New paradigms, opportunities, and challenges on transceiver design[J]. IEEE Wireless Communications, 2020, 27(2): 180-187.

\bibitem{ref72}
Tang W, Dai J Y, Chen M Z, et al. MIMO transmission through reconfigurable intelligent surface: System design, analysis, and implementation[J]. IEEE Journal on Selected Areas in Communications, 2020, 38(11): 2683-2699.

\bibitem{ref73}
Tan X, Sun Z, Koutsonikolas D, et al. Enabling indoor mobile millimeter-wave networks based on smart reflect-arrays[C]//IEEE INFOCOM 2018-IEEE Conference on Computer Communications. Piscataway: IEEE Press, 2018: 270-278.

\bibitem{ref74}
Yildirim I, Uyrus A, Basar E. Modeling and analysis of reconfigurable intelligent surfaces for indoor and outdoor applications in future wireless networks[J]. IEEE Transactions on Communications, 2021, 69(2): 1290-1301.

\bibitem{ref75}
Perovi$\acute{\rm c}$ N S, Di Renzo M, Flanagan M F. Channel capacity optimization using reconfigurable intelligent surfaces in indoor mmWave environments[C]//2020-2020 IEEE International Conference on Communications. Piscataway: IEEE Press, , 2020: 1-7.

\bibitem{ref76}
Pan C, Ren H, Wang K, et al. Multicell MIMO communications relying on intelligent reflecting surfaces[J]. IEEE Transactions on Wireless Communications, 2020, 19(8): 5218-5233.

\bibitem{ref77}
Cui M, Zhang G, Zhang R. Secure wireless communication via intelligent reflecting surface[J]. IEEE Wireless Communications Letters, 2019, 8(5): 1410-1414.

\bibitem{ref78}
Yu X, Xu D, Sun Y, et al. Robust and secure wireless communications via intelligent reflecting surfaces[J]. IEEE Journal on Selected Areas in Communications, 2020, 38(11): 2637-2652.

\bibitem{ref79}
Xu P, Chen G, Pan G, et al. Ergodic Secrecy Rate of RIS-Assisted Communication Systems in the Presence of Discrete Phase Shifts and Multiple Eavesdroppers[J]. IEEE Wireless Communications Letters, 2020: 1-1.

\bibitem{ref80}
$\ddot{\rm O}$zdogan $\ddot{\rm O}$, Bj$\ddot{\rm o}$rnson E, Larsson E G. Using intelligent reflecting surfaces for rank improvement in MIMO communications[C]//IEEE International Conference on Acoustics, Speech and Signal Processing. Piscataway: IEEE Press, , 2020: 9160-9164.

\bibitem{ref81}
Kisseleff S, Martins W A, Al-Hraishawi H, et al. Reconfigurable Intelligent Surfaces for Smart Cities: Research Challenges and Opportunities[J]. IEEE Open Journal of the Communications Society, 2020, 1: 1781-1797.

\bibitem{ref82}
Wu Q, Zhang S, Zheng B, et al. Intelligent reflecting surface aided wireless communications: A tutorial[J]. arXiv preprint arXiv:2007.02759, 2020.

\bibitem{ref83}
Yang L, Meng F, Zhang J, et al. On the performance of RIS-assisted dual-hop UAV communication systems[J]. IEEE Transactions on Vehicular Technology, 2020, 69(9): 10385-10390.

\bibitem{ref84}
Ma D, Ding M, Hassan M. Enhancing cellular communications for UAVs via intelligent reflective surface[C]//IEEE Wireless Communications and Networking Conference. Piscataway: IEEE Press, 2020: 1-6.

\bibitem{ref85}
Li S, Duo B, Yuan X, et al. Reconfigurable intelligent surface assisted UAV communication: Joint trajectory design and passive beamforming[J]. IEEE wireless communications letters, 2020, 9(5): 716-720.

\bibitem{ref86}
Ge L, Dong P, Zhang H, et al. Joint beamforming and trajectory optimization for intelligent reflecting surfaces-assisted UAV communications[J]. IEEE Access, 2020, 8: 78702-78712.

\bibitem{ref87}
Zhang Q, Saad W, Bennis M. Reflections in the sky: Millimeter wave communication with UAV-carried intelligent reflectors[C]//IEEE Global Communications Conference. Piscataway: IEEE Press, 2019: 1-6.

\bibitem{ref88}
Lu H, Zeng Y, Jin S, et al. Enabling panoramic full-angle reflection via aerial intelligent reflecting surface[C]//IEEE International Conference on Communications Workshops. Piscataway: IEEE Press, 2020: 1-6.

\bibitem{ref89}
Liu X, Liu Y, Chen Y. Machine learning empowered trajectory and passive beamforming design in UAV-RIS wireless networks[J]. IEEE Journal on Selected Areas in Communications, 2020: 1-1.

\bibitem{ref90}
Rotshild D, Abramovich A. Wideband reconfigurable entire Ku-band metasurface beam-steerable reflector for satellite communications[J]. IET Microwaves, Antennas \& Propagation, 2018, 13(3): 334-339.

\bibitem{ref91}
Zhou Z, Ge N, Liu W, et al. RIS-aided offshore communications with adaptive beamforming and service time allocation[C]//IEEE International Conference on Communications. Piscataway: IEEE Press, 2020: 1-6.

\bibitem{ref92}
Wu Q, Zhang R. Weighted sum power maximization for intelligent reflecting surface aided SWIPT[J]. IEEE Wireless Communications Letters, 2019, 9(5): 586-590.

\bibitem{ref93}
Wang Z, Shi Y, Zhou Y, et al. Wireless-powered over-the-air computation in intelligent reflecting surface aided IoT networks[J]. IEEE Internet of Things Journal, 2020, 8(3): 1-1.

\bibitem{ref94}
Yu G, Chen X, Zhong C, et al. Design, Analysis, and Optimization of a Large Intelligent Reflecting Surface-Aided B5G Cellular Internet of Things[J]. IEEE Internet of Things Journal, 2020, 7(9): 8902-8916.

\bibitem{ref95}
Gacanin H, Di Renzo M. Wireless 2.0: Towards an intelligent radio environment empowered by reconfigurable meta-surfaces and artificial intelligence[J]. arXiv preprint arXiv:2002.11040, 2020.

\bibitem{ref96}
Zappone A, Di Renzo M, Debbah M. Wireless networks design in the era of deep learning: Model-based, AI-based, or both?[J]. IEEE Transactions on Communications, 2019, 67(10): 7331-7376.

\bibitem{ref97}
Gao J, Zhong C, Chen X, et al. Unsupervised learning for passive beamforming[J]. IEEE Communications Letters, 2020, 24(5): 1052-1056.

\bibitem{ref98}
Liaskos C, Tsioliaridou A, Nie S, et al. An interpretable neural network for configuring programmable wireless environments[C]//Proceedings of the IEEE 20th International Workshop on Signal Processing Advances in Wireless Communications. Piscataway: IEEE Press, 2019: 1-5.

\bibitem{ref99}
Huang C, Alexandropoulos G C, Yuen C, et al. Indoor signal focusing with deep learning designed reconfigurable intelligent surfaces[C]//Proceedings of the IEEE 20th International Workshop on Signal Processing Advances in Wireless Communications. Piscataway: IEEE Press, 2019: 1-5.

\bibitem{ref100}
Feng K, Wang Q, Li X, et al. Deep reinforcement learning based intelligent reflecting surface optimization for MISO communication systems[J]. IEEE Wireless Communications Letters, 2020, 9(5): 745-749.

\bibitem{ref101}
Shan T, Pan X, Li M, et al. Coding programmable metasurfaces based on deep learning techniques[J]. IEEE Journal on Emerging and Selected Topics in Circuits and Systems, 2020, 10(1): 114-125.

\bibitem{ref102}
Huang C, Mo R, Yuen C. Reconfigurable intelligent surface assisted multiuser MISO systems exploiting deep reinforcement learning[J]. IEEE Journal on Selected Areas in Communications, 2020, 38(8): 1839-1850.

\bibitem{ref103}
Song H, Zhang M, Gao J, et al. Unsupervised Learning based Joint Active and Passive Beamforming Design for Recongurable Intelligent Surfaces Aided Wireless Networks[J]. IEEE Communications Letters, 2020.

\bibitem{ref104}
Taha A, Zhang Y, Mismar F B, et al. Deep reinforcement learning for intelligent reflecting surfaces: Towards standalone operation[C]//Proceedings of the IEEE 21st International Workshop on Signal Processing Advances in Wireless Communications . Piscataway: IEEE Press, 2020: 1-5.

\bibitem{ref105}
Khan S, Khan K S, Haider N, et al. Deep-learning-aided detection for reconfigurable intelligent surfaces[J]. arXiv preprint arXiv:1910.09136, 2019.

\bibitem{ref106}
Taha A, Alrabeiah M, Alkhateeb A. Enabling large intelligent surfaces with compressive sensing and deep learning[J]. arXiv preprint arXiv:1904.10136, 2019.

\bibitem{ref107}
Elbir A M, Papazafeiropoulos A, Kourtessis P, et al. Deep channel learning for large intelligent surfaces aided mm-wave massive MIMO systems[J]. IEEE Wireless Communications Letters, 2020, 9(9): 1447-1451.

\bibitem{ref108}
Liu X, Liu Y, Chen Y, et al. RIS enhanced massive non-orthogonal multiple access networks: Deployment and passive beamforming design[J]. IEEE Journal on Selected Areas in Communications, 2020.

\bibitem{ref109}
Lee B, Lee G Y, Hong J Y. Metasurface Devices for AR/VR[C]//Proceedings of the Conference on Lasers and Electro-Optics. Piscataway: IEEE Press, 2019: 1-2.

\bibitem{ref110}
Genevet P, Capasso F. Holographic optical metasurfaces: a review of current progress[J]. Reports on Progress in Physics, 2015, 78(2): 024401.

\bibitem{ref111}
Lee G Y, Hong J Y, Hwang S H, et al. Metasurface eyepiece for augmented reality[J]. Nature communications, 2018, 9(1): 1-10.

\bibitem{ref112}
Lee G Y, Yoon G, Lee S Y, et al. Complete amplitude and phase control of light using broadband holographic metasurfaces[J]. Nanoscale, 2018, 10(9): 4237-4245.

\bibitem{ref113}
Li L, Cui T J, Ji W, et al. Electromagnetic reprogrammable coding-metasurface holograms[J]. Nature communications, 2017, 8(1): 1-7.

\bibitem{ref114}
Guo L, Wang Z, Shen L, et al. Metasurface holograms with arbitrary phase control of electromagnetic wavefront[C]//Proceedings of the IEEE International Conference on Computational Electromagnetics. Piscataway: IEEE Press, 2018: 1-2.

\bibitem{ref115}
Ding X, Wang Z, Guan C, et al. Spatial Rotation Operations on Huygens Metasurface Hologram in Microwave Regime[J]. IEEE Transactions on Magnetics, 2019, 55(10): 1-4.

\bibitem{ref116}
Smy T J, Stewart S A, Gupta S. Surface susceptibility synthesis of metasurface holograms for creating electromagnetic illusions[J]. IEEE Access, 2020, 8: 93408-93425.

\bibitem{ref117}
Haimov T, Aydin K, Scheuer J. Reconfigurable Holograms Using VO 2-Based Tunable Metasurface[J]. IEEE Journal of Selected Topics in Quantum Electronics, 2020, 27(1): 1-8.

\bibitem{ref118}
Chaccour C, Soorki M N, Saad W, et al. Risk-based optimization of virtual reality over terahertz reconfigurable intelligent surfaces[C]//Proceedings of the IEEE International Conference on Communications. Piscataway: IEEE Press, 2020: 1-6.

\bibitem{ref119}
Wu Q, Zhang R. Beamforming optimization for intelligent reflecting surface with discrete phase shifts[C]//Proceedings of the IEEE International Conference on Acoustics, Speech and Signal Processing. Piscataway: IEEE Press,2019: 7830-7833.

\bibitem{ref120}
Saeed T, Skitsas C, Kouzapas D, et al. Fault adaptive routing in metasurface controller networks[C]//Proceedings of the 11th International Workshop on Network on Chip Architectures . Piscataway: IEEE Press, 2018: 1-6.

\bibitem{ref121}
Kouvaros P, Kouzapas D, Philippou A, et al. Formal verification of a programmable hypersurface[C]//Proceedings of the International Workshop on Formal Methods for Industrial Critical Systems. Piscataway: Springer Press, Cham, 2018: 83-97.

\bibitem{ref122}
Petrou L, Karousios P, Georgiou J. Asynchronous circuits as an enabler of scalable and programmable metasurfaces[C]//Proceedings of the IEEE International Symposium on Circuits and Systems. Piscataway: IEEE Press, 2018: 1-5.

\bibitem{ref123}
Abadal S, Mestres A, Torrellas J, et al. Medium access control in wireless network-on-chip: A context analysis[J]. IEEE Communications Magazine, 2018, 56(6): 172-178.

\bibitem{ref124}
Bj$\ddot{\rm o}$rnson E, Sanguinetti L. Demystifying the power scaling law of intelligent reflecting surfaces and metasurfaces[C]//Proceedings of the IEEE 8th International Workshop on Computational Advances in Multi-Sensor Adaptive Processing. Piscataway: IEEE Press, 2019: 549-553.

\bibitem{ref125}
Shen H, Xu W, Gong S, et al. Secrecy rate maximization for intelligent reflecting surface assisted multi-antenna communications[J]. IEEE Communications Letters, 2019, 23(9): 1488-1492.

\bibitem{ref126}
Guan X, Wu Q, Zhang R. Intelligent reflecting surface assisted secrecy communication: Is artificial noise helpful or not?[J]. IEEE Wireless Communications Letters, 2020, 9(6): 778-782.

\bibitem{ref127}
Mishra D, Johansson H. Channel estimation and low-complexity beamforming design for passive intelligent surface assisted MISO wireless energy transfer[C]//Proceedings of the IEEE International Conference on Acoustics, Speech and Signal Processing. Piscataway: IEEE Press, 2019: 4659-4663.

\bibitem{ref128}
Zuo J, Liu Y, Qin Z, et al. Resource allocation in intelligent reflecting surface assisted NOMA systems[J]. IEEE Transactions on Communications, 2020, 68(11): 7170-7183.

\end{thebibliography}
\end{document}